\documentclass[a4paper,fleqn]{cas-dc}
\usepackage[numbers,square,sort&compress]{natbib}
\usepackage{xcolor}
\usepackage{tcolorbox}
\usepackage[section]{placeins}
\usepackage{float}

\usepackage[table]{xcolor}
\usepackage{colortbl}
\definecolor{hlprimary}{RGB}{255,235,205}
\definecolor{hlsecondary}{RGB}{220,235,250}
\definecolor{hlalert}{RGB}{255,220,220}
\definecolor{hlmuted}{RGB}{240,240,240}
\usepackage{rotating}

\usepackage[table]{xcolor}
\usepackage{colortbl}
\usepackage{tikz}
\usepackage{subcaption}
\usepackage{threeparttable}

\definecolor{hg1}{RGB}{237,247,237}
\definecolor{hg2}{RGB}{210,235,210}
\definecolor{hg3}{RGB}{170,215,170}
\definecolor{hg4}{RGB}{120,190,120}
\definecolor{hg5}{RGB}{ 70,160, 70}

\definecolor{hb1}{RGB}{237,242,250}
\definecolor{hb2}{RGB}{206,221,242}
\definecolor{hb3}{RGB}{156,184,224}
\definecolor{hb4}{RGB}{ 99,138,196}
\definecolor{hb5}{RGB}{ 45, 92,165}

\usepackage{tikz}
\definecolor{mbblue}{RGB}{99,138,196}
\definecolor{mbgray}{RGB}{237,242,250}
\newcommand{\sfbar}[2]{%
  \begin{tikzpicture}[baseline=-0.5ex]
    \useasboundingbox (0,0) rectangle (30pt, 2.4ex);
    \fill[mbgray, rounded corners=0.4pt] (0,0) rectangle (30pt, 2.4ex);
    \fill[mbblue, rounded corners=0.4pt] (0,0) rectangle (#1, 2.4ex);
    \node[font=\footnotesize, text=black!85] at (15pt, 1.2ex) {#2};
  \end{tikzpicture}%
}

\def\tsc#1{\csdef{#1}{\textsc{\lowercase{#1}}\xspace}}
\tsc{WGM}
\tsc{QE}

\begin{document}
\let\WriteBookmarks\relax
\def\floatpagepagefraction{1}
\def\textpagefraction{.001}

\shorttitle{}    

\shortauthors{}  

\title [mode = title]{When Passing Tests Hides Vulnerabilities: An Empirical Study of Silent Failures in Agentic Systems}  

\tnotemark[1] 
\tnotetext[1]{} 

\author[1]{Wenji Bai}[orcid=0000-0002-3283-3483]
\cormark[1]
\ead{wenji.bai@tuni.fi}
\credit{Conceptualization, Methodology, Software, Formal analysis, Investigation, Data curation, Writing - original draft, Visualization}

\author[1]{Muhammad Waseem}[orcid=0000-0001-7488-2577]
\ead{muhammad.waseem@tuni.fi}
\credit{Conceptualization, Methodology, Validation, Review, Editing, Supervision, Project administration}

\author[1]{Zeeshan Rasheed}[orcid=0000-0001-9655-3096]
\ead{zeeshan.rasheed@tuni.fi}
\credit{Formal analysis, Investigation}

\author[1]{Jaakko Peltonen}[orcid=0000-0003-3485-8585]
\ead{jaakko.peltonen@tuni.fi}
\credit{Validation, Review, Editing, Supervision}

\author[1]{Pekka Abrahamsson}[orcid=0000-0002-4360-2226]
\ead{pekka.abrahamsson@tuni.fi}
\credit{Resources, Supervision, Review, Editing}

\affiliation[1]{organization={Faculty of Information Technology and Communication Sciences, Tampere University},
city={Tampere},
country={Finland}}

\cortext[1]{Corresponding author}

\fntext[1]{}

\begin{abstract}
LLM-based agents for automated code repair have received significant attention in recent years from both research and software engineering practice perspectives. However, limited attention has been paid to patches that pass syntactic and functional verification but still retain or introduce security vulnerabilities. The aim of this research is to systematically identify and categorize such silent failures in LLM-based agentic code repair.
 
We conducted an empirical study using 1,030 valid execution traces produced by seven agent frameworks with GPT-4o-mini across two security-focused datasets, SecurityEval and CVEfixes. Through three iterations of qualitative coding and manual verification, 170 confirmed silent failures were identified. The key results are: (i) Three main categories of silent failures were identified: Omission, Introduction, and Inadequacy. Omission accounts for 48.2\% of the confirmed failures, Introduction for 30.6\%, and Inadequacy for 21.2\%. (ii) Ten fine-grained failure codes were classified under these three categories, showing how agents omit required security controls, apply incomplete defenses, or introduce new vulnerabilities during repair. (iii) Current test-passing evaluation and LLM-based reviewer roles were insufficient to expose or intercept these failures in the confirmed cases. (iv) Similar insecure solutions appeared across different frameworks, suggesting possible shared model-, prompt-, or task-level influences, while single-agent and multi-agent systems showed different failure profiles.
 
The results of this study will assist researchers and practitioners in improving the evaluation of LLM-based agentic code repair and developing targeted verification methods that go beyond functional correctness and cover all generated artifacts.
\end{abstract}




\maketitle

\section{Introduction}\label{sec:introduction}
Silent failures in automated software repair refer to patches that pass functional validation but still contain flaws, in particular security vulnerabilities. In automated program repair, a patch that compiles and passes all available test cases is conventionally termed \emph{plausible}, but plausible patches are not necessarily correct~\citep{qi2015analysis}. 
We extend this concept into the security domain, and
to systematically make the distinction explicit, we adopt a four-level verification hierarchy: syntactic validity (L0), functional validation (L1), static security analysis (L2), and exploitability assessment (L3). Thus, a patch may compile without errors (L0) and pass unit tests (L1), while still containing a vulnerability detectable through static security analysis (L2) or exploitability testing (L3) \citep{perry2023users, peng2025cweval}. Unlike overt errors such as compilation failures or test breakages, silent failures evade standard quality gates and can remain undetected and reach production systems.
 
This issue has become increasingly relevant with the rise of LLM-based agentic systems. Recent tools can handle multiple steps of the repair workflow, from fault diagnosis to patch implementation, marking a shift from earlier code completion assistants~\citep{yang2024swe,chen2024coder, bouzenia2025repairagent}. Furthermore, multi-agent frameworks organize multiple LLMs into role-specialized pipelines that mirror human development teams~\citep{qian2024chatdev,hong2023metagpt,wu2024autogen}. These architectures use role specialization and interaction to improve output quality. However, recent studies report that multi-agent systems do not consistently outperform single-agent approaches, and their failure modes are not well characterized~\citep{xia2025agentless,cemri2026why}. These systems are increasingly deployed for security-sensitive tasks, including vulnerability patching and authentication repair~\citep{zhang2024autocoderover}, which increases the risk of undetected vulnerabilities.
 
Current evaluation practices can reinforce this risk. Benchmarks such as SWE-bench evaluate success primarily through test passing~\citep{jimenez2024swebench,chen2021evaluating}. In Continuous Integration and Continuous Delivery (CI/CD) pipelines, results are typically reported based solely on test passing
criteria. When patches pass, reviewers may reasonably assume that agents have produced functionally correct repairs. However, an agent-generated patch can satisfy a narrowly scoped test case while retaining an exploitable vulnerability. For instance, an agent might resolve a data processing error to pass a functional test, yet continue to process untrusted user input without proper sanitization, a well-known vulnerability pattern~\citep{chen2025red,wang2025vulnrepaireval}.

Prior work has explored related issues, but a systematic account of silent failures remains absent. Existing taxonomies categorize observable failure modes in multi-agent systems, yet they rely on explicit easily noticeable signals like system crashes or task non-completion; failures that produce no such obvious signals fall outside their scope~\citep{cemri2026why}. Similarly, current attribution frameworks offer mechanisms for identifying fault-inducing steps but still assume that failures are externally visible~\citep{zhang2025which, zhang2026agentracer}. In automated program repair, recent work has advocated for multi-dimensional evaluation incorporating security benchmarks and static analysis, yet these approaches have not been applied to characterize failures within multi-agent repair pipelines~\citep{kuzmina2025spring,tihanyi2025vulnerability}.
 
Motivated by this gap, we formulate the following three research questions: 
 
\begin{itemize}
    \item \textit{RQ1: What types of silent failures occur in LLM-based agent code repair, and how prevalent are they?}
 
    \textit{Rationale:} Existing taxonomies classify observable failure modes in multi-agent systems, but they primarily address explicit indications such as crashes, errors, or task non-completion. Silent failures, defined as patches that pass functional verification yet contain security deficiencies, remain insufficiently characterized. This question develops an empirically grounded taxonomy of silent failure types and quantifies their prevalence across frameworks and datasets, thereby establishing the basis for the subsequent vulnerability-context analyses.
 
    \item \textit{RQ2: How do silent failures propagate through agent pipelines, and what role do architectural differences play?}

    \textit{Rationale:} Multi-agent frameworks employ specialized roles such as planners, engineers, and reviewers, but it remains unclear whether these role structures intercept security deficiencies or allow them to propagate across pipeline stages. This question traces silent failures through agent roles and pipeline stages, examines whether reviewer roles detect them, and compares propagation patterns between multi-agent and single-agent architectures. The goal is to clarify how architectural structure shapes the emergence, propagation, and interception of silent failures.
 
    \item \textit{RQ3: Which vulnerability types and code contexts are most susceptible to silent failures, and what severity and exploitability levels do they exhibit?}
 
    \textit{Rationale:} Vulnerability characteristics may shape both the form silent failures take and their severity and exploitability, yet this relationship has received limited empirical attention. This question examines how silent failures distribute across CWE families, code locations, severity levels, and exploitability scores. The goal is to identify cross-dimensional patterns that can support targeted verification strategies rather than treating silent failures as a homogeneous defect class.
\end{itemize}

To address these research questions, we conduct an empirical study of silent failures across seven LLM-based agent frameworks and two security-focused datasets. Our analysis encompasses 1,030 execution traces, from which we identify 170 confirmed silent failures. Through iterative qualitative coding of execution traces and related artifacts, we develop an empirically grounded taxonomy centered on \emph{security-compliance decoupling} from mere unit-test passing. Specifically, this paper makes the following contributions:
 
\begin{itemize}
    \item \textbf{An empirically grounded taxonomy of silent failures.} We propose an empirically grounded taxonomy that classifies silent failures along four dimensions (failure type, originator role, injection stage, and code location), extending beyond existing failure classifications by explicitly modeling failures that evade standard detection mechanisms.
 
    \item \textbf{A cross-framework empirical analysis.} We analyze 1,030 execution traces from seven agent frameworks across two security-focused datasets, providing quantitative evidence on the prevalence, distribution, and characteristics of silent failures.
 
    \item \textbf{Failure propagation analysis across agent roles.} By tracing silent failures through multi-agent execution logs, we identify how security omissions introduced at one pipeline stage propagate through subsequent stages, and compare propagation patterns between multi-agent and single-agent systems.
 
   \item \textbf{A public dataset.} We publicly release the study dataset online \citep{Bai2026dataset}, including execution traces, generated patches, verification results, and taxonomy codes, to enable researchers and practitioners to access, replicate, and validate our findings.
\end{itemize}

The remainder of this paper is organized as follows.  Section~\ref{sec:methodology} describes our research methodology. Section~\ref{sec:findings} presents the empirical results. Section~\ref{sec:discussion} discusses implications, and Section~\ref{sec:threats} addresses threats to validity. Section~\ref{sec:related} reviews related work. Section~\ref{sec:conclusion} concludes with directions for future work.

\section{Research Methodology}
\label{sec:methodology}
This section describes the research methodology used to identify, verify, and analyze silent failures in LLM-based agentic code repair. We first present the overall research design, followed by the dataset and agent framework selection criteria. We then describe the experimental setup, the multi-level verification framework, the manual review process, and the qualitative coding and reliability procedures used to construct and validate the taxonomy.

\subsection{Research Design Overview}
\label{sec:design_overview}

\begin{figure*}[!t]
\centering
\includegraphics[width=\textwidth]{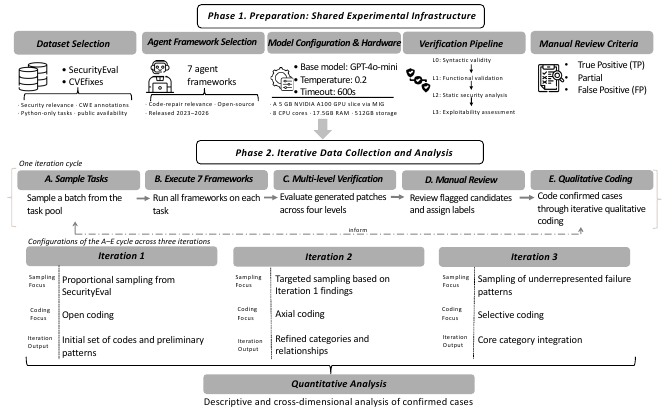}
\caption{Overview of the research methodology.}
\label{fig:overview}
\end{figure*}

The study consists of two phases, as illustrated in Figure~\ref{fig:overview}: (i) preparation of the shared experimental infrastructure, and (ii) iterative data collection and analysis. This design supports our three research questions on failure categorization, propagation mechanisms, and vulnerability susceptibility, while enabling consistent comparison across seven agent frameworks. The first phase covers dataset selection, framework selection, model configuration, the multi-level verification pipeline, and manual review criteria. The second phase repeats the same procedure across three iterations: sampling tasks, executing all seven frameworks, evaluating generated patches through L0--L3 verification, manually reviewing flagged candidates, and qualitatively coding the confirmed cases. Observations from each iteration informed the sampling and analytical focus of the next.
\subsection{Dataset Selection}
\label{sec:datasets}
 
\begin{table}[htbp]
\centering
\small
\caption{Summary of selected datasets. \emph{Pool} denotes the eligible task count after language and format filtering; \emph{Sampled} denotes the number of tasks drawn across three iterations.}
\label{tab:datasets}
\begin{tabular}{lcc>{\bfseries}r}
\hline
 & SecurityEval & CVEfixes & Total \\
\hline
Task Type & Code generation & Vulnerability repair & --\\
Original  & 130             & 5,365                & 5,495 \\
Pool      & 121             & 1,115                & 1,236 \\
Sampled   & 85              & 65                   & 150 \\
Labels    & CWE             & CVE/CWE              & --\\
\hline
\end{tabular}
\end{table}

To construct our dataset pool, we first defined four selection criteria: (i) security relevance, (ii) the availability of Common Weakness Enumeration (CWE) annotations to enable verification against expected vulnerability types, (iii) uniform execution conditions, requiring Python-only tasks so that all seven frameworks could be evaluated under identical conditions, and (iv) public availability. Based on these criteria, we selected two security-focused datasets, which together provided a combined pool of 1,236 Python tasks. Table~\ref{tab:datasets} summarizes the selected datasets, which are described below.

\begin{itemize}
    \item \textbf{SecurityEval}~\citep{siddiq2022securityeval} provides 130 predefined code generation prompts across 69 CWE categories. We retained the 121 Python tasks and excluded non-Python entries. Each task presents an insecure code context and requires the agent to generate a secure version, serving as a benchmark for initial pattern discovery and cross-framework comparison.
    \item \textbf{CVEfixes}~\citep{bhandari2021cvefixes} contains 5,365 real-world vulnerability fixes linking Common Vulnerabilities and Exposures (CVE) records to code changes in open-source projects. We filtered for Python single-file fixes with CWE annotations, yielding 1,115 tasks. CVEfixes tasks involve patching known vulnerabilities in existing production code and often require adding entirely new security logic rather than merely substituting an insecure API, representing real-world repair conditions.
\end{itemize}

\subsection{Agent Framework Selection}
\label{sec:frameworks}

To select the agent frameworks for this study, we first defined three criteria: (i) relevance to code repair, requiring that each framework be applicable to vulnerability repair tasks either by design or through reasonable adaptation; (ii) open-source availability; and (iii) recency, requiring that each framework has been released between 2023 and 2026 to reflect recent advances in the field. Based on these criteria, we systematically selected seven agent frameworks representing distinct coordination patterns. Table~\ref{tab:frameworks} provides an overview of the selected frameworks and their coordination architectures.

\begin{table*}[htbp]
\centering
\caption{Selected agent frameworks and their coordination architectures.}
\label{tab:frameworks}
\begin{tabular}{lllp{7cm}}
\hline
Framework & Type & Architecture & Selection Rationale \\
\hline
\rowcolor{gray!15}
\multicolumn{4}{l}{\textit{Multi-Agent Systems}} \\
MetaGPT \citep{hong2023metagpt}    & Multi & Assembly Line & Sequential role handoffs enable stage-level failure tracing \\
ChatDev \citep{qian2024chatdev}     & Multi & Hierarchical Workflow & Phase-based workflow with explicit role transitions; validates pattern generalization across pipeline architectures \\
AutoGen \citep{wu2024autogen}     & Multi & Dynamic Conversation & Flexible orchestration with runtime role assignment; captures emergent failure patterns in less structured collaboration \\
AgentCoder \citep{huang2023agentcoder}  & Multi & Test-driven Loop & Independent test designer and coder roles; tests whether agent-generated tests mask vulnerabilities \\
\hline
\rowcolor{gray!15}
\multicolumn{4}{l}{\textit{Single-Agent Systems}} \\
Mini SWE-Agent (MiniSWE) \citep{yang2024swe} & Single & ReAct Loop & Lightweight single-agent baseline; isolates coordination effects by removing inter-agent communication \\
Aider \citep{gauthier2024aider}        & Single & Edit-feedback Loop & Structured diff output with lint/test feedback loops; represents developer-tool-style interaction \\
OpenHands \citep{wang2024openhands}    & Single & Edit-feedback Loop & CLI-based sandbox with file editing and shell execution in a single generalist loop
	 \\
\hline
\end{tabular}
\end{table*}

\subsection{Experimental Setup}
\label{sec:exp_setup}
This section outlines the model and prompt configuration, hardware platform, four-level verification framework, and manual review and confirmation criteria used in our study. In this setup, each task is independently processed by all frameworks to produce generated patches under a standardized configuration; every patch is then evaluated through a four-level verification pipeline (L0 to L3); finally, patches flagged at L2 or L3 are manually reviewed and confirmed as True Positive, Partial, or False Positive.

\subsubsection{Model Configuration and Hardware}
\label{sec:model_config}
To reduce variability due to model differences, all frameworks used GPT-4o-mini (temperature 0.2) and were executed in isolated environments with a shared verification toolchain. We chose this model to balance response quality, cost, and scalability across the evaluation of seven frameworks. Experiments were run on a 5 GB NVIDIA A100 MIG slice with 8 CPU cores, 17.5 GB RAM, and 512 GB storage, with a 600-second timeout.

To maintain comparability across frameworks, we standardized the prompts by keeping the task content identical while adapting the format to each framework's input schema. For example, ChatDev takes the task as a single project-level requirement string passed to its task parameter, while MiniSWE takes the same content as a user message inside a ReAct loop; in both cases, the core task description and code context remain identical, and only the input schema differs. At the dataset level, SecurityEval prompts included the original insecure code context, whereas CVEfixes prompts paired the vulnerable code with a repair instruction specifying the target CVE/CWE. Beyond these dataset-specific differences, no additional security prompts were injected unless the task itself explicitly implied a security requirement.

\subsubsection{Multi-Level Verification Framework}
\label{sec:verification}

As outlined in Section~\ref{sec:introduction}, passing functional tests does not guarantee security correctness \citep{qi2015analysis, dolcetti2026helping}. We therefore evaluate every generated patch through a four-level verification pipeline:

\begin{itemize}
    \item \textbf{Level~0 (L0), Syntactic validity.} The generated code parses without syntax errors.
 
    \item \textbf{Level~1 (L1), Functional validation.} The generated code passes available functional checks, including import resolution, structural correctness verification, and execution of associated test cases.
 
    \item \textbf{Level~2 (L2), Static security analysis.} The generated code is analyzed with Bandit (a tool to find Python code security issues, maintained by the Python Code Quality Authority). We run Bandit and map its findings to CWE identifiers using a predefined lookup table. A generated patch fails L2 if it contains any HIGH-severity finding, where severity is assigned based on CWE type.
 
    \item \textbf{Level~3 (L3), Exploitability assessment.} To support exploitability-oriented candidate screening, we apply a heuristic scoring mechanism that estimates risk based on the co-occurrence of user input sources and dangerous sinks, weighted by CWE severity \citep{perry2023users}. Concretely, the score is computed from four signal groups extracted by pattern matching: (i) user-input sources, such as external inputs received from users, requests, or command-line arguments; (ii) dangerous sinks, such as dynamic execution, shell invocation, unsafe deserialization, or security-sensitive database operations, each assigned a base weight reflecting its exploitation severity; (iii) indicators that the original vulnerability pattern targeted by the task is still present; and (iv) context-aware indicators capturing risky co-occurrences, such as a hard-coded credential near authentication logic. The total score increases when a user-input source and a dangerous sink co-occur in the same patch, and is further increased when the sink matches the target CWE of the task. The score is reduced when safe coding practices are present. In particular, bounded discounts are applied for safeguards such as input validation and sanitization, parameterized query construction, certificate or host verification, and secure replacement of known unsafe operations. The exploitability score $R$ is computed as follows, clipped to the range $[0,100]$:
    
    \begin{equation}
\label{eq:l3-score}
R = \textstyle\sum_{s\in\mathcal{I}} w_s + \sum_{j\in\mathcal{K}} b_j\,c\,t_j + 10\,\delta_{\text{vuln}} - \min(2n_{\text{safe}},\,10),
\end{equation}
where $\mathcal{I}$ and $\mathcal{K}$ are the input-source types and dangerous-sink matches detected in the patch, weighted by $w_s$ and $b_j$ respectively; $c = 1.5$ when an input source co-occurs with a sink and $1.0$ otherwise; $t_j = 1.3$ when the sink CWE matches the task target CWE and $1.0$ otherwise; $\delta_{\text{vuln}} = 1$ when the original target-CWE pattern is still present; and $n_{\text{safe}} \in \{0,\dots,5\}$ is the number of safe-practice categories detected. The weights $w_s$ follow three tiers by controllability (web request data 4, interactive and standard input 3, environment or file reads 2), and the weights $b_j$ span 6 to 12 by exploitation severity, with dynamic code execution highest (12), command execution next (10), template rendering (9) and unsafe deserialization (8) in the middle, and formatted-string operations lowest (7 and 6). For patches spanning multiple files, $R$ is taken as the maximum across files.

A generated patch whose total score reaches the threshold of 15 is treated as a silent failure candidate. We set the threshold to 15 and the associated signal weights in Eq.~\eqref{eq:l3-score} based on calibration with Iteration 1 candidates. This configuration retained the high-risk source-sink co-occurrences observed during the initial calibration while reducing isolated low-severity patterns. Since all candidates undergo manual review before confirmation, the threshold serves as a sensitivity parameter for candidate selection rather than a final classification boundary.
\end{itemize}

We define patches that fail L0 or L1 as \emph{explicit failures}, since their deficiencies are visible through standard quality gates. Patches that pass both levels but are flagged at L2 or L3 are treated as \emph{silent failure candidates}: on the surface, they look like successful repairs because they are syntactically valid and pass available functional checks, yet later security-oriented checks indicate that they may still contain or introduce a vulnerability. More formally, this is defined as:

\begin{equation}
\mathit{SF}_{\text{candidate}} = (L_0^{\text{pass}} \wedge L_1^{\text{pass}}) \wedge
(L_2^{\text{fail}} \vee L_3^{\text{fail}})
\label{eq:sf_condition}
\end{equation}

In other words, a patch is considered a silent failure candidate if it appears valid and passes available functional checks, but still raises concerns in later security checks.

As an illustrative case, consider a SecurityEval command-injection task that required repairing code that built and executed a shell command from external input. The agent-generated patch added a length check and renamed surrounding variables, but retained the original shell-execution call. The patch passed L0 and L1 because it parsed successfully and the available tests used only benign inputs that produced the expected output. However, L2 still flagged the reachable unsafe shell execution, and L3 assigned a high score because a user-controlled source co-occurred with a dangerous sink matching the task CWE. Manual review therefore confirmed the patch as a silent failure: it appeared successful under functional evaluation but preserved the original injection path.

\subsubsection{Manual Review and Confirmation Criteria}
\label{sec:manual_review}
The L2/L3 pipeline flags candidates automatically, but static analysis can produce irrelevant or non-actionable warnings. The first author conducted the initial manual review of every flagged case following a structured coding procedure. To reduce individual bias, a subset of cases was independently reviewed by another author as part of the reliability assessment described in Section~\ref{sec:reliability}.

\paragraph{Review Process and Consistency Mechanisms}
To maintain consistency across the entire dataset, we applied an iterative review protocol:
\begin{itemize}
    \item[(i)] \textbf{Standardized Annotation:} The first author examined the generated patch, L2/L3 findings, and original task description for each case. Judgments and metadata (e.g., failure type, severity) were recorded using a codebook \citep{macqueen1998codebook}, which strictly defined the admissible values for each dimension.
    
    \item[(ii)] \textbf{Iterative Alignment (Within-iteration):} The review was conducted iteration-by-iteration. At the end of each iteration, iteration-level summaries were compiled to analyze the distribution of judgments and re-examine borderline cases.
    
    \item[(iii)] \textbf{Retrospective Revision (Across iterations):} Following iteration 3, all codes were consolidated to ensure uniformity. If a newly observed case prompted a refinement to the coding scheme, earlier judgments were retrospectively updated. Adjustments were recorded during the coding process to support transparency and consistency.
    
    \item[(iv)] \textbf{Independent Validation:} To assess inter-rater reliability, another author independently recoded 20\% of the corpus (reliability measures reported in Section \ref{sec:reliability}).
\end{itemize}
\begin{table*}[htbp]
\centering
\caption{Iteration-based data collection plan.}
\label{tab:iterations}
\begin{tabular}{lrllll}
\hline
Iteration & Tasks & Source & Sampling Logic & Coding Activity \\
\hline
Iteration 1 & 50 & SecurityEval & Proportional by CWE & Open Coding \\
Iteration 2 & 50 & SecurityEval + CVEfixes & Targeted framework--dataset pairs & Axial Coding \\
Iteration 3 & 50 & SecurityEval + CVEfixes & Underrepresented patterns & Selective Coding \\
\hline
\multicolumn{2}{l}{Total: 150 tasks} & \multicolumn{3}{l}{$\times$ 7 agents = 1,050 target traces} \\
\hline
\end{tabular}
\end{table*}
\paragraph{Category Definitions}
Based on the review, each candidate was assigned to exactly one of the following categories:
\begin{itemize}
    \item \textbf{True Positive (TP):} The patch contains a clear instance of the target vulnerability. The agent either made no attempt to fix it or applied a superficial fix that entirely failed to address the root cause. For example, validating input length while leaving an unsafe call intact.
    
    \item \textbf{Partial:} The patch contains a visible and relevant mitigation attempt, but the defense remains incomplete and a plausible exploit path is still present in the code. Unlike TP, a defense against the target CWE is visibly attempted; unlike FP, the defense does not fully neutralize the weakness. For instance, the patch blocks one specific bypass while leaving the sink reachable through another path, or constrains a symptom rather than the underlying sink.
    
    \item \textbf{False Positive (FP):} The patch does not contain a genuine security weakness relevant to the task, even though it was flagged by the pipeline. This typically occurs when the flagged pattern appears only in non-deployable test code, when an effective safeguard already neutralizes the risky operation, or when the static tool matches code that appears dangerous syntactically but is harmless in context.
\end{itemize}

Both TP and Partial cases are retained in the confirmed corpus. We include Partial cases because they exhibit the same code-level failure patterns as TPs and are relevant for taxonomy construction, which focuses on what the agent did wrong rather than on exploitability severity. In our corpus, Partial cases account for 39 of 170 confirmed failures.
\begin{table*}[t]
\centering
\caption{Inter-rater agreement across coding dimensions.}
\label{tab:kappa_agreement}
\scriptsize
\setlength{\tabcolsep}{3pt}

\begin{subtable}[t]{0.28\textwidth}
\centering
\caption{D0 triage verdict}
\label{tab:kappa_d0}
\begin{tabular}{lccc|c}
\toprule
R1 $\backslash$ R2 & TP & Partial & FP & Total \\
\midrule
TP      & 19 & 2 & 2 & 23 \\
Partial & 0  & 3 & 1 & 4  \\
FP      & 1  & 1 & 6 & 8  \\
\midrule
Total   & 20 & 6 & 9 & 35 \\
\bottomrule
\end{tabular}

\end{subtable}
\hfill
\begin{subtable}[t]{0.31\textwidth}
\centering
\caption{D1 axial category}
\label{tab:kappa_d1}
\begin{tabular}{lcccc|c}
\toprule
R1 $\backslash$ R2 & Om. & Inad. & Intro. & FP & Total \\
\midrule
Om.    & 8 & 0 & 0 & 1 & 9 \\
Inad.  & 0 & 7 & 0 & 2 & 9 \\
Intro. & 0 & 2 & 7 & 0 & 9 \\
FP     & 1 & 1 & 0 & 0 & 2 \\
\midrule
Total  & 9 & 10 & 7 & 3 & 29 \\
\bottomrule
\end{tabular}

\end{subtable}
\hfill
\begin{subtable}[t]{0.38\textwidth}
\centering
\caption{Composite D0$\times$D1}
\label{tab:kappa_composite}
\begin{tabular}{lcccccc|c}
\toprule
R1 $\backslash$ R2 & FP & P-I & P-O & T-I & T-Intro & T-O & Total \\
\midrule
FP      & 6 & 1 & 0 & 0 & 0 & 1 & 8 \\
P-I     & 1 & 2 & 0 & 0 & 0 & 0 & 3 \\
P-O     & 0 & 0 & 1 & 0 & 0 & 0 & 1 \\
T-I     & 1 & 1 & 0 & 4 & 0 & 0 & 6 \\
T-Intro & 0 & 1 & 0 & 1 & 7 & 0 & 9 \\
T-O     & 1 & 0 & 0 & 0 & 0 & 7 & 8 \\
\midrule
Total   & 9 & 5 & 1 & 5 & 7 & 8 & 35 \\
\bottomrule
\end{tabular}

\end{subtable}

\vspace{4pt}
\footnotesize
\raggedright
Note: Om. = Omission; Inad. = Inadequacy; Intro. = Introduction; P-I = Partial-Inadequacy; P-O = Partial-Omission; T-I = TP-Inadequacy; T-Intro = TP-Introduction; T-O = TP-Omission. 
\end{table*}
\subsection{Data Collection and Analysis}
\label{sec:coding}

\subsubsection{Data Collection}
\label{sec:sampling}
Data collection consisted of two steps: sampling and preprocessing.

\begin{enumerate}
    \item  \textbf{Sampling}. We purposively sampled 150 tasks in three iterations of 50 each \citep{patton2014qualitative}, drawn without replacement from a pool of 1,236 eligible tasks. In this study, an iteration refers to one sampling round followed by agent execution and preliminary verification-based screening. Each iteration involved: (i) selecting a batch of tasks; (ii) executing the selected agent frameworks on these tasks; (iii) normalizing the outputs into the common trace schema; (iv) screening the resulting traces through the Multi-Level Verification Framework; and (v) reviewing the resulting candidate cases to guide the sampling focus of the next round \citep{corbin2014basics, stol2016grounded}. The three iterations served different purposes. (i) Iteration~1 sampled from SecurityEval proportionally across CWE categories to obtain broad initial coverage. (ii) Iteration~2 used targeted sampling based on the patterns observed in Iteration~1, especially recurring or ambiguous failure patterns that required further evidence. (iii) Iteration~3 focused on underrepresented open codes and boundary cases to supplement missing failure patterns and refine category coverage. This iterative strategy was intended to support progressive coverage of silent-failure patterns rather than statistical representativeness. Table~\ref{tab:iterations} summarizes the sampling details.
    
    \item \textbf{Preprocessing}. To address the inconsistencies in execution trace formats across different frameworks, we normalized all data into a unified schema to enable cross-framework comparison. This step was informed by recent multi-agent failure dataset construction practice, where cross-framework trace analysis requires standardized representations to support consistent annotation and comparison \citep{cemri2026why}. After normalization, we excluded 20 traces that did not produce valid outputs due to timeouts, runtime crashes, or incomplete generation, leaving 1,030 valid traces. We then evaluated each valid trace through the multi-level verification framework described in Section \ref{sec:verification}. Generated patches that failed at L0 or L1 were classified as explicit failures. Those that passed both L0 and L1 but were flagged at L2 or L3 yielded 252 silent failure candidates. After manual review, 170 cases were confirmed as silent failures. These 170 cases form the corpus for the coding procedure described next.
\end{enumerate}
\subsubsection{Coding Procedure}
\label{sec:coding_procedure}
\begin{figure*}[htbp]
\centering
\includegraphics[width=\textwidth]{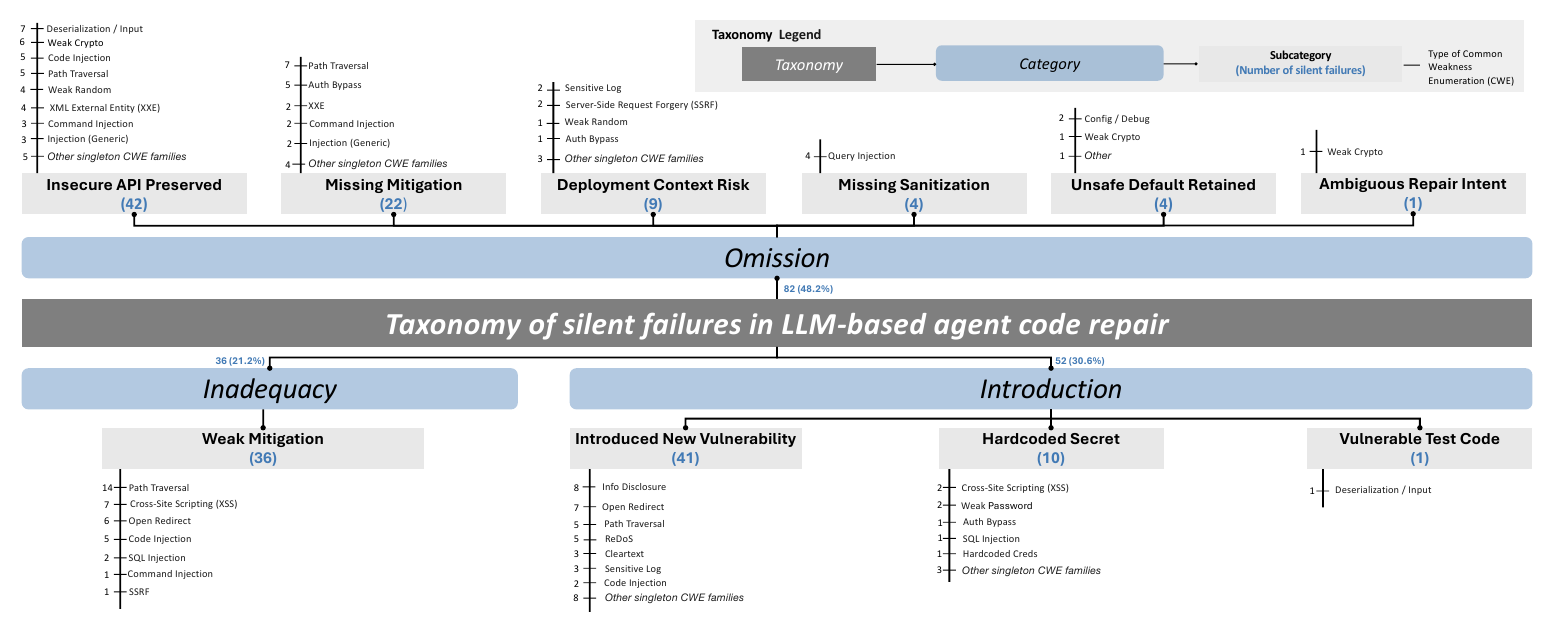}
\caption{Taxonomy of silent failures in LLM-based agent code repair. The taxonomy classifies failure mechanisms rather than CWE families; therefore, the same CWE family may appear under different open codes when it arises through different mechanisms. ``Other singleton CWE families'' denotes one-off CWE families grouped only for visualization to avoid over-fragmenting the figure.}
\label{fig:taxonomy}
\end{figure*}
We analyzed confirmed cases using an iterative thematic coding process inspired by Grounded Theory \citep{glaser2017discovery}. Coding was conducted in three stages, using individual execution traces (one framework’s output per task) as the unit of analysis.

\begin{enumerate}
    \item  \textit{Open coding}. The first author examined each confirmed case, including the generated code, L2/L3 findings, and the original task context, and assigned descriptive labels to observed failure patterns. No codes were predefined. Each case was also annotated with a propagation chain, a severity rating reflecting the coder's overall assessment of the weakness (Critical, High, Medium, Low), and candidate detection mechanisms.
    \item  \textit{Axial coding}. The first author analyzed the initial codes and organized them along four dimensions: failure type, originator role, injection stage, and code location. Relationships between dimensions were examined and documented.
    \item  \textit{Selective coding}. The first author reviewed and refined the coding results. We identified an overarching theme, security-compliance decoupling, that connects the categories into a unified explanation of silent failures. Categories were merged, split, or dropped based on discussion among all authors.
\end{enumerate}

Throughout this process, each newly coded case was compared against earlier cases in the same category to check for consistency. We considered coding complete when no new codes emerged in two consecutive batches of 50 traces~\citep{corbin2014basics}.
\subsubsection{Reliability Measures}
\label{sec:reliability}

To ensure the consistency of the qualitative coding process and mitigate individual bias, we performed an Inter-Rater Reliability (IRR) analysis using Cohen's Kappa~\citep{cohen1960coefficient}. A random subset of 35 cases was independently coded by another author (R2), in parallel with the primary coder (R1). Each case was assessed along two coding dimensions: (i)~\textit{D0 (triage verdict)}, classifying the case as TP, PARTIAL, or FP; and (ii)~\textit{D1 (axial category)}, recording the axial category assigned by a rater, with FP retained as an exclusion label for cases a rater did not treat as a silent failure.

Agreement was evaluated at three levels, as shown in Table~\ref{tab:kappa_agreement}. First, we measured agreement on the D0 triage verdict, which distinguishes true positives, partial cases, and false positives. Second, we measured agreement on the D1 axial category, which assigns non-false-positive cases to the main silent-failure categories. This analysis included 29 cases because cases coded as FP by both raters were excluded from the axial-category calculation. Third, we computed a composite D0$\times$D1 agreement score, which combines the triage decision and the axial category into a single coding decision. We report the composite score as the primary inter-rater reliability metric because it captures both whether a candidate was treated as a confirmed silent failure and how it was categorized.

The primary composite agreement was $\kappa=0.71$, indicating substantial agreement. The D0 triage verdict and D1 axial category also reached substantial agreement, with $\kappa=0.63$ and $\kappa=0.66$, respectively. All disagreements were resolved through discussion among the authors. These discussions were used to clarify ambiguous coding rules, refine category boundaries, and update the final codebook before applying it consistently to the full corpus. In addition, each newly coded case was compared against prior cases in the same category to reduce category drift during the iterative coding process.

\section{Results}
\label{sec:findings}
\begin{figure*}[htbp]
  \centering
  \includegraphics[width=\textwidth]{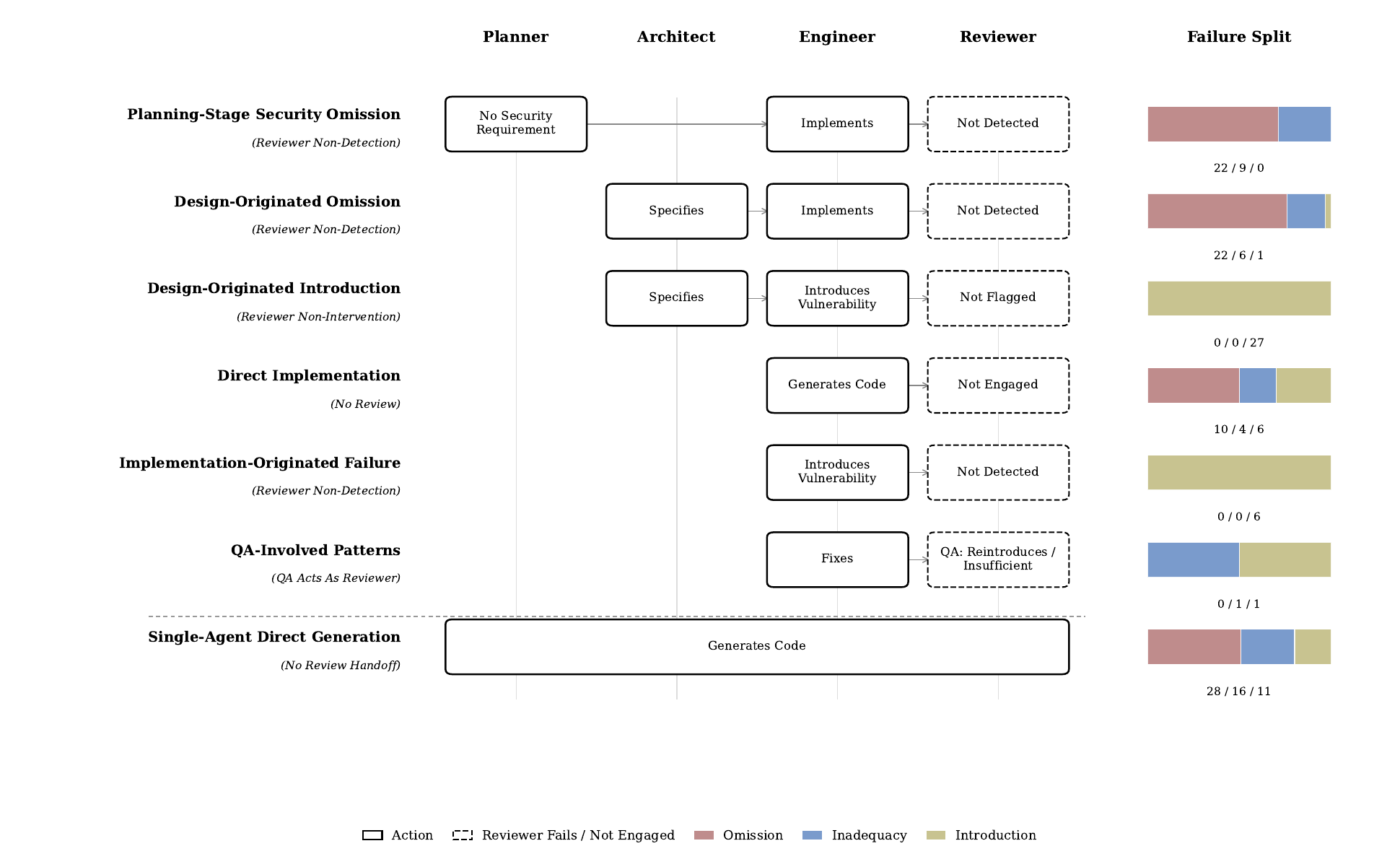}
  \caption{Propagation chain patterns of silent failures by origin, failure mechanism, and review outcome. Color-coded bars on the right show the axial category split for each pattern.}
  \label{fig:chains}
\end{figure*}

\begin{table*}[htbp]
\centering
\caption{Distribution and exploitability profile of silent failure open codes.}
\label{tab:taxonomy_distribution}
\small
\setlength{\tabcolsep}{4pt}
\begin{tabular}{llp{6cm}rcr}
\hline
\textbf{Axial Category} & \textbf{Open Code} & \textbf{Description} & \textbf{$n$} & \textbf{\% in Total} & \textbf{Mean L3} \\
\hline
Omission & Insecure API Preserved & Dangerous API retained without secure substitution & 42 & \sfbar{30.00pt}{24.7} & 16.7 \\
 & Missing Mitigation & Required security control entirely absent & 22 & \sfbar{15.67pt}{12.9} & 5.8 \\
 & Deployment Context Risk & Security adequacy depends on deployment context & 9 & \sfbar{6.44pt}{5.3} & 4.2 \\
 & Missing Sanitization & Input sanitization absent at a critical point & 4 & \sfbar{2.91pt}{2.4} & 2.0 \\
 & Unsafe Default Retained & Insecure configuration value left unchanged & 4 & \sfbar{2.91pt}{2.4} & 1.0 \\
 & Ambiguous Repair Intent & Repair intent ambiguous; security effect indeterminate & 1 & \sfbar{0.73pt}{0.6} & 0.0 \\
 & \textit{Subtotal} & & \textit{82} & \textit{48.2} & \textit{10.7} \\
\hline
Inadequacy & Weak Mitigation & Defense attempted but bypassable or incomplete & 36 & \sfbar{25.75pt}{21.2} & 17.7 \\
 & \textit{Subtotal} & & \textit{36} & \textit{21.2} & \textit{17.7} \\
\hline
Introduction & Introduced New Vulnerability & Repair introduced a vulnerability not originally present & 41 & \sfbar{29.27pt}{24.1} & 6.9 \\
 & Hardcoded Secret & Cryptographic secret embedded as string literal & 10 & \sfbar{7.17pt}{5.9} & 10.4 \\
 & Vulnerable Test Code & Test code reintroduced a fixed vulnerability & 1 & \sfbar{0.73pt}{0.6} & 29.0 \\
 & \textit{Subtotal} & & \textit{52} & \textit{30.6} & \textit{8.0} \\
\hline
\textbf{Total} & & & \textbf{170} & \textbf{100.0} & \textbf{11.4} \\
\hline
\end{tabular}
\end{table*}
This section presents the results of our empirical investigation, addressing the three research questions outlined in Section~\ref{sec:introduction}: RQ1 on failure taxonomy and prevalence, RQ2 on propagation patterns across agent architectures, and RQ3 on the distribution of silent failures across vulnerability types, code locations, and severity or exploitability levels. Throughout this section, we present axial categories in \textbf{boldface}, open codes in Title Case (e.g., Insecure API Preserved), and conceptual terms in \emph{italics}. We report the silent failure taxonomy and prevalence in Section~\ref{sec:rq1} (see Figure~\ref{fig:taxonomy} and Table~\ref{tab:taxonomy_distribution}), propagation patterns across agent architectures in Section~\ref{sec:rq2} (see Table~\ref{tab:chains} and Figure~\ref{fig:chains}), and the distribution of silent failures across vulnerability types, code locations, and severity or exploitability levels in Section~\ref{sec:rq3} (see Figure~\ref{fig:rq3} and Table~\ref{tab:d4_axial}).

\subsection{Silent Failure (RQ1)}
\label{sec:rq1}

Figure~\ref{fig:taxonomy} provides the complete taxonomy of silent failures identified in this study. The taxonomy was derived through qualitative coding of 170 confirmed silent failures and consists of three axial categories: Omission, Inadequacy, and Introduction with ten open codes. Table~\ref{tab:taxonomy_distribution} complements the figure by reporting the quantitative distribution and mean L3 exploitability risk score for each open code and category.

\begin{figure*}[htbp]
  \centering
  \includegraphics[width=\textwidth]{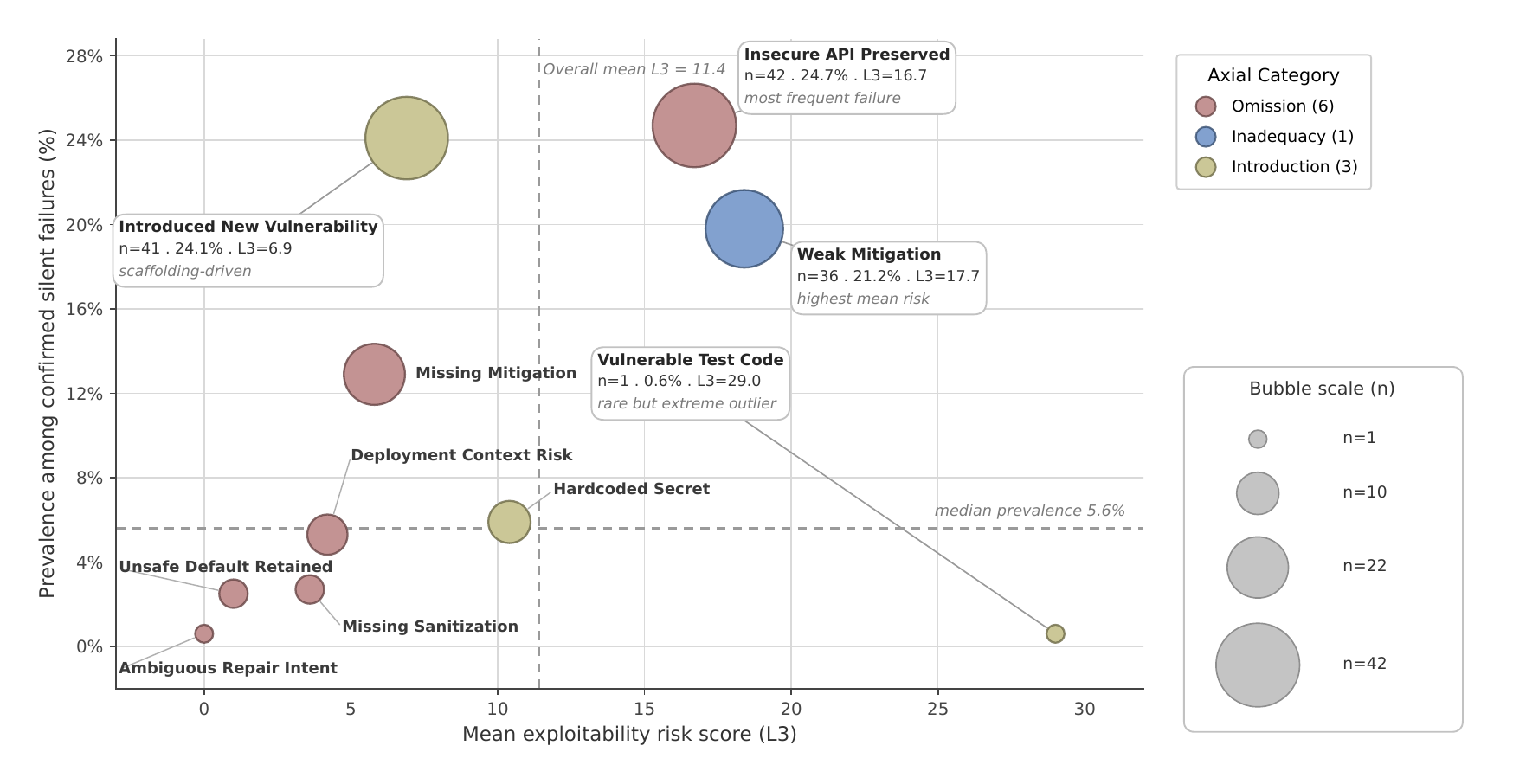}
  \caption{Risk-prevalence landscape of silent failure codes. Each bubble represents one open code; horizontal axis shows mean L3 exploitability risk, vertical axis shows prevalence (\%) among the $n=170$ confirmed silent failures, bubble size encodes the absolute count $n$, and color encodes the axial category. Reference lines mark the overall mean L3 ($11.4$) and median prevalence across codes.}
  \label{fig:taxonomy_dist}
\end{figure*}

The results show that Omission is the most frequently observed category (82/170, 48.2\%), followed by Introduction (52/170, 30.6\%) and Inadequacy (36/170, 21.2\%). A silent failure may exhibit more than one security symptom. However, we assign each case to the dominant failure mechanism, namely the mechanism most directly responsible for the failure to satisfy the security intent of the task. Figure~\ref{fig:taxonomy_dist} further shows how the open codes differ in both prevalence and exploitability risk.

\textbf{1. Omission (82/170, 48.2\%).} Omission refers to cases in which the agent fails to implement a necessary security measure despite the task context indicating that such a measure is required. This is the largest category in the taxonomy of silent failures. We identified and classified six open codes under Omission (see 
Table~\ref{tab:taxonomy_distribution}).
Each of them is briefly described below.
 
\textbullet\ Insecure API Preserved (42, 24.7\%): The agent retains a dangerous API call from the input context without substituting a secure alternative. This is the dominant code within Omission and the most frequent code in the entire corpus. We observed that agents often applied cosmetic adjustments such as renaming variables or adding length checks, while leaving the vulnerable call intact.

\textbullet\ Missing Mitigation (22, 12.9\%): The agent omits an entire security control that the task requires but the existing code does not suggest. This code appeared predominantly in CVEfixes tasks, which more frequently require adding new security logic rather than substituting an API. Unlike Insecure API Preserved, which reflects insecure pattern reproduction, Missing Mitigation captures cases where the required defense must be inferred from task intent rather than copied from the input context.

\textbullet\ Deployment Context Risk (9, 5.3\%): The generated code may appear secure in isolation, but its security adequacy depends on deployment-time conditions that the task does not specify. The agent leaves these conditions unaddressed, transferring the security burden to downstream operators. Examples include hardcoded permissive CORS policies and unverified trust assumptions about upstream input sources.

\textbullet\ Missing Sanitization (4, 2.4\%): Input sanitization is absent at a critical point along the data flow. This code differs from Missing Mitigation in that the required defense is specifically input validation or escaping at an identifiable sink. The small count reflects how seldom this isolated pattern appeared without overlapping with other Omission codes.

\textbullet\ Unsafe Default Retained (4, 2.4\%): An insecure configuration value present in the input is left unchanged in the generated code. Observed instances include retained debug flags and default credentials in authentication-related tasks.

\textbullet\ Ambiguous Repair Intent (1, 0.6\%): The repair intent is ambiguous and the security effect cannot be determined from the generated artifact alone. We retained this code to mark a boundary case rather than discard the trace, but it represents an edge case in our corpus.
 
\textbf{2. Inadequacy (36/170, 21.2\%).} Inadequacy refers to cases in which the agent appears to recognize the security requirement and attempts a defense, but the resulting mechanism remains incomplete or bypassable. This category differs from Omission in that a security-relevant action is present in the generated code; what is missing is its sufficiency. We identified a single open code under Inadequacy. It is briefly described below.

\textbullet\ Weak Mitigation (36, 21.2\%): The agent implements a defense that is partial, bypassable, or built on an incorrect security primitive. Among the open codes occurring in more than one case, Weak Mitigation carries the highest mean L3 exploitability risk (17.7), suggesting that partially implemented defenses are associated with a viable attack surface in the confirmed cases. Cases concentrated in vulnerability types such as path traversal, command injection, and eval-based execution. For example, one agent implemented path traversal prevention using a string-level comparison function rather than a path-aware alternative; another added a command allowlist but retained an injection-prone execution mode; a third constructed a restricted evaluation sandbox whose filtering rules were too permissive to prevent known escape techniques.

\textbf{3. Introduction (52/170, 30.6\%).} Introduction refers to cases in which the repair process itself injects a vulnerability that was not present in the original code. Unlike Omission and Inadequacy, which describe what the agent failed to do, Introduction describes a new defect arising from agent action. We identified and classified three open codes under Introduction. Each of them is briefly described below.

\textbullet\ Introduced New Vulnerability (41, 24.1\%): The agent generates code that introduces a vulnerability absent from the original input. This code is the dominant pattern within Introduction (41/52, 78.8\%). Multi-agent frameworks contributed a higher share, concentrated in ChatDev and MetaGPT. These frameworks also generated additional artifacts, such as configuration files, build scripts, and test scaffolding, where some of these introduced vulnerabilities appeared.

\textbullet\ Hardcoded Secret (10, 5.9\%): The agent embeds cryptographic secrets (e.g., signing keys, tokens, passwords) as string literals in the generated code. The surrounding authentication or session logic is often otherwise correct, which makes this pattern particularly easy to overlook in functional review.

\textbullet\ Vulnerable Test Code (1, 0.6\%): The agent introduces a vulnerability through generated test code rather than production code. Although rare, this case stands out with an L3 score of 29.0, far above the Introduction category mean of 8.0, since the test artifact reintroduced a previously fixed vulnerability into the project workspace.

Taken together, the three axial categories share a common characteristic: in every confirmed case, the agent produced code that satisfied L0 and L1 verification criteria yet failed to achieve the security intent of the task. We refer to this recurring pattern as \emph{security-compliance decoupling}: the observed disconnect between an agent's externally observable compliance signals and the actual security properties of the generated artifact. The three categories correspond to distinct points at which this disconnect appears: Omission involves a gap between the task and appropriate API selection; Inadequacy involves a gap between identifying the security requirement and implementing a complete defense; Introduction involves a gap between addressing the primary task and managing the security context of all generated artifacts.

\begin{tcolorbox}[colback=gray!5, colframe=gray!88, title=Key Findings from RQ1]
We identified three recurring mechanisms of silent failures: Omission, Introduction, and Inadequacy. These mechanisms share a common pattern of \emph{security-compliance decoupling}, where generated patches appear to satisfy functional requirements but fail to preserve security properties. We further derived ten open codes to capture the specific failure patterns, with Insecure API Preserved and Introduced New Vulnerability being the most frequent. Among the three mechanisms, Inadequacy showed the highest mean exploitability risk, suggesting that partially implemented defenses may be especially dangerous because they create an appearance of security while remaining bypassable.
\end{tcolorbox}
 \begin{figure*}[!htbp]
\centering
\includegraphics[width=\textwidth]{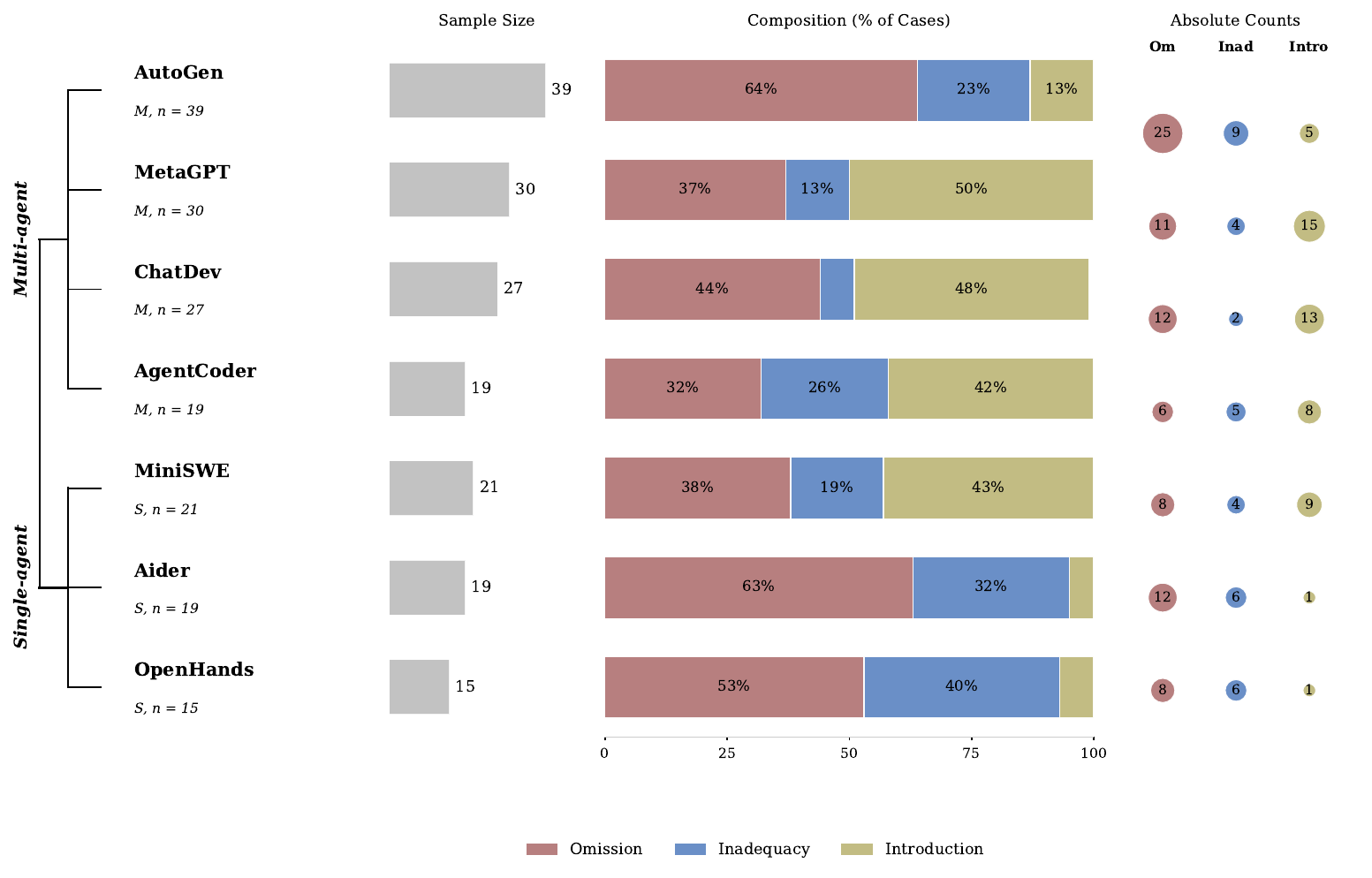}
\caption{Distribution of silent failure categories across agent frameworks.}
\label{fig:rq2}
\end{figure*}
\subsection{Propagation Patterns Across Agent Architectures (RQ2)}
\label{sec:rq2}
We use the same three axial categories as in Section~\ref{sec:rq1}: Omission, Inadequacy, and Introduction, denoting absent defenses, incomplete defenses, and newly introduced vulnerabilities, respectively. The propagation patterns are summarized in Table~\ref{tab:chains}, and Figure~\ref{fig:chains} illustrates the operational chain of each pattern by origin stage, failure mechanism, and review outcome. The architectural comparison is reported in Table~\ref{tab:arch_comparison}, and the distribution of categories across frameworks is shown in Figure~\ref{fig:rq2}. Each aspect is discussed below.

Originator attribution was assigned during manual coding (see Section~\ref{sec:coding}). For each confirmed case, coders inspected the per-role trace outputs and identified the earliest role whose output made the security deficiency observable in the trajectory. When the deficiency first appeared in a planning or design artifact, the corresponding Planner or Architect role was coded as the originator; when it first appeared in generated code, the Engineer role was coded as the originator. Disagreements were resolved through discussion. For instance, in one path traversal task, the Planner decomposed the request into file-reading and response-formatting steps but omitted the need to constrain user-provided paths to a safe base directory. The Engineer implemented the underspecified plan by joining user input with a base path without canonicalization, and the Reviewer approved the output after checking functional completeness rather than path-handling security. Because the first observable deficiency appeared in the planning artifact and persisted through implementation and review, we coded the Planner as the originator and the propagation pattern as a planning-stage omission cascade.

Among multi-agent systems, the three most frequent patterns all involve a Reviewer role that does not intercept the security deficiency. The most common is the planning-stage omission cascade, in which the Planner decomposes the task without including security requirements, the Engineer implements code consistent with the underspecified plan, and the Reviewer does not flag the issue. Following closely is design-originated omission, where the Architect specifies a design that the Engineer implements and the Reviewer evaluates without detecting the deficiency. The third is design-originated introduction; ChatDev and MetaGPT accounted for all instances of this pattern, as their pipelines produce configuration templates and scaffolding that expand the attack surface.

A Reviewer role was active in most multi-agent cases. In the confirmed cases we reviewed, the Reviewer rarely flagged the relevant security deficiency before it reached the final output. This pattern appeared across the multi-agent frameworks in our corpus, despite their differing coordination architectures.

\begin{table}[htbp]
\centering
\caption{Propagation chain patterns of silent failures.}
\label{tab:chains}
\small
\setlength{\tabcolsep}{6pt}
\renewcommand{\arraystretch}{1.2}
\begin{tabular}{lrrr}
\hline
\textbf{Propagation Pattern} & \textbf{Om} & \textbf{Inad} & \textbf{Intro} \\
\hline
\rowcolor{gray!15}
\multicolumn{4}{l}{\textit{Multi-Agent} } \\
Planning-Stage Omission Cascade                       & 22 & 9  & 0  \\
Design-Originated Omission                            & 22 & 6  & 1  \\
Design-Originated Introduction                        & 0  & 0  & 27 \\
Direct Implementation (No Review)                     & 10 & 4  & 6  \\
QA-Involved Patterns                                  & 0  & 1  & 1  \\
Implementation-Originated                             & 0  & 0  & 6  \\
\hline
\rowcolor{gray!15}
\multicolumn{4}{l}{\textit{Single-Agent}} \\
Single-Agent Direct Generation                        & 28 & 16 & 11 \\
\hline
\textbf{Total}                                        & \textbf{82} & \textbf{36} & \textbf{52} \\
\hline
\end{tabular}
\end{table}

Beyond these multi-agent patterns, Table~\ref{tab:arch_comparison} shows distinct failure distributions between the two architectural families. For instance, Introduction failures are more common in multi-agent systems (41/115, 35.7\%) than in single-agent ones (11/55, 20.0\%), reflecting the additional configuration and scaffolding artifacts their pipelines produce. By contrast, Inadequacy accounts for a higher share of single-agent failures (16/55, 29.1\%) than of multi-agent ones (20/115, 17.4\%). This indicates that the observed failure profile differs between the two architecture groups: Introduction is more frequent in multi-agent systems, whereas Inadequacy represents a larger share of single-agent failures. However, the proportion of Omission is nearly identical across architectures (54/115, 47.0\% and 28/55, 50.9\% respectively).

\begin{table}[htbp]
\centering
\caption{Axial category distribution by architecture type.}
\label{tab:arch_comparison}
\small
\setlength{\tabcolsep}{3pt}
\begin{tabular}{lrrrr}
\hline
  & Om. & Inad. & Intro. & Total \\
\hline
Multi-agent  & 54 (47.0\%) & 20 (17.4\%) & 41 (35.7\%) & 115 \\
Single-agent & 28 (50.9\%) & 16 (29.1\%) & 11 (20.0\%) & 55 \\
\hline
Total        & 82          & 36          & 52           & 170 \\
\hline
\end{tabular}
\end{table}
Further, we observed task-level convergence across frameworks. 
Specifically, on 26 tasks, three or more frameworks each produced a silent failure that shared the same confirmed CWE; these comparable failures account for 62.9\% (107/170) of all
confirmed cases. We treated failures as comparable when they occurred on the same task and shared the same confirmed CWE. Four of the tasks were particularly convergent in that six of the seven frameworks produced a silent failure with the same CWE. This shows that a subset of tasks elicited similar silent-failure patterns across otherwise distinct frameworks, rather than isolated framework-specific failures. However, because all experiments used the same base model and related task prompts, this convergence should not be attributed to framework architecture alone. The present design does not isolate model effects from other shared factors, such as task design, prompt content, or dataset-specific vulnerability patterns. We therefore report cross-framework convergence as an empirical pattern in our corpus.
\begin{figure*}[htbp]
\centering
\includegraphics[width=1.0\textwidth]{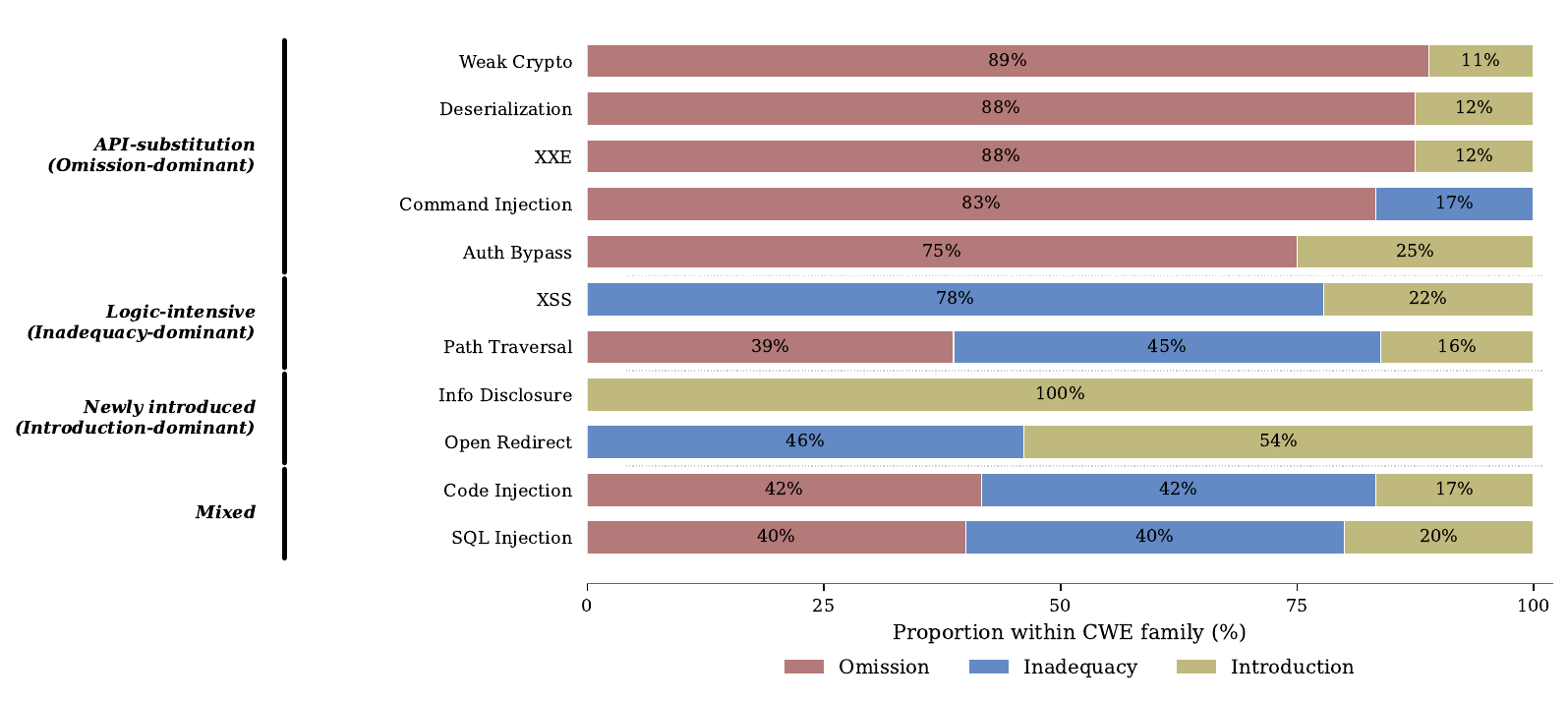}
\caption{Silent failure category distribution by CWE family.}
\label{fig:rq3}
\end{figure*}
\begin{tcolorbox}[colback=gray!5, colframe=gray!88, title=Key Findings from RQ2]
We observed that silent failures propagate through three primary multi-agent mechanisms: planning-stage security omission cascades, design-originated omissions, and design-originated introductions. In the multi-agent cases we reviewed, the Reviewer rarely flagged a security deficiency before it reached the final output. We further observed different failure profiles across agent architectures: multi-agent systems produced more \textsc{Introduction} failures, consistent with their additional artifact generation, whereas single-agent systems produced more \textsc{Inadequacy} failures, consistent with attempted but incomplete defenses. Finally, similar failures recurred across frameworks on the same tasks, indicating cross-framework convergence within the observed corpus.
\end{tcolorbox}

\subsection{Distribution Across Vulnerability Types, Code Locations, and Risk Levels (RQ3)}
\label{sec:rq3}

The distribution across CWE families is shown in Figure~\ref{fig:rq3}, the distribution across code locations is reported in Table~\ref{tab:d4_axial}, and the severity and risk profiles are reported in Table~\ref{tab:severity_axial}. These three views capture, respectively, what kind of vulnerability is affected, where in the code the failure resides, and how severe and exploitable it is. Across all three, the axial categories align with distinct profiles rather than spreading evenly, as detailed below.

Specifically, as Figure~\ref{fig:rq3} illustrates, Omission accounts for the large majority of cases in families where the fix involves substituting a known insecure API with a secure alternative: weak cryptography (8/9, 88.9\%), deserialization (7/8, 87.5\%), XXE (7/8, 87.5\%), and command injection (5/6, 83.3\%). These families share a common trait: the insecure pattern is visible in the code, and its secure counterpart is well-documented. Conversely, Inadequacy clusters in families whose fix must cover multiple input cases or bypass routes, including XSS and path traversal. Finally, Introduction is associated with newly created flaws, such as information disclosure and open redirect.
 
Besides vulnerability family, code location provides a complementary view of where silent failures reside. Table~\ref{tab:d4_axial} presents the distribution across code locations. Omission is most frequent in API calls and input handling, corresponding to cases where the agent retained a dangerous function call or omitted required validation logic. Inadequacy is predominantly located in input handling (33/36), reflecting cases where agents attempted validation or sanitization but implemented it incorrectly. Introduction is highly concentrated in configuration and business logic, corresponding to vulnerabilities added through scaffolding generation or auxiliary functionality. The alignment between code location and axial category suggests that the taxonomy dimensions capture relatively distinct aspects of the failure phenomenon.

\begin{table}[htbp]
\centering
\caption{Silent failure distribution by code location and axial category.}
\label{tab:d4_axial}
\small
\setlength{\tabcolsep}{4pt}
\begin{tabular}{lrrrrc}
\hline
\textbf{Code Location} & \textbf{Om.} & \textbf{Inad.} & \textbf{Intro.} & \textbf{Total} & \textbf{In Total} \\
\hline
Input Handling & 26 & 33 & 2 & 61 & \sfbar{30.00pt}{35.9} \\
Configuration & 5 & 0 & 32 & 37 & \sfbar{18.20pt}{21.8} \\
API Call & 35 & 0 & 0 & 35 & \sfbar{17.21pt}{20.6} \\
Business Logic & 7 & 3 & 9 & 19 & \sfbar{9.34pt}{11.2} \\
Authentication & 2 & 0 & 7 & 9 & \sfbar{4.43pt}{5.3} \\
Other & 7 & 0 & 2 & 9 & \sfbar{4.43pt}{5.3} \\
\hline
\textbf{Total} & \textbf{82} & \textbf{36} & \textbf{52} & \textbf{170} & \textbf{100.0\%} \\
\hline
\end{tabular}
\vspace{0.3em}

\footnotesize
\textit{Note.} The \textit{In Total} column shows each code location's share of the 170-case corpus, with bar length proportional to that share. Three locations (Input handling, Configuration, and API call) together account for over three-quarters of all confirmed silent failures, and each is the dominant location for one axial category respectively (Inadequacy, Introduction, and Omission).
\end{table}
 
The severity and exploitability profiles further differentiate the three categories. Severity follows the rating assigned during the coding procedure (see Section~\ref{sec:coding_procedure}), and the L3 risk score is the exploitability estimate produced by our L3 procedure (see Section~\ref{sec:verification}). With these definitions, Table~\ref{tab:severity_axial} reports the severity distribution and L3 risk by category. Omission contains the largest number of critical-severity cases, predominantly involving unrestricted code execution and insecure deserialization. Introduction has the lowest severity profile, with most cases rated as medium (46/52, 88.5\%), reflecting that introduced vulnerabilities tend to be configuration-level issues rather than direct execution paths. However, Inadequacy shows the highest mean L3 risk (17.7), above Omission (10.7) and Introduction (8.0). Many Inadequacy cases occur in logic-intensive families such as XSS and path traversal, where partial defenses can leave bypass routes open.

\begin{table}[htbp]
\centering
\caption{Severity distribution and L3 risk by axial category.}
\label{tab:severity_axial}
\small
\setlength{\tabcolsep}{6pt}
\begin{tabular}{lrrrr}
\hline
 & \textbf{Om.} & \textbf{Inad.} & \textbf{Intro.} & \textbf{Total} \\
\hline
\multicolumn{5}{l}{\textit{Severity}} \\
\quad Critical & \cellcolor{hb4}17 & \cellcolor{hb3}5  & \cellcolor{hb1}0  & 22 \\
\quad High     & \cellcolor{hb5}31 & \cellcolor{hb4}17 & \cellcolor{hb3}5  & 53 \\
\quad Medium   & \cellcolor{hb5}34 & \cellcolor{hb4}14 & \cellcolor{hb5}46 & 94 \\
\quad Low      & \cellcolor{hb1}0  & \cellcolor{hb1}0  & \cellcolor{hb2}1  & 1 \\
\hline
\multicolumn{5}{l}{\textit{L3 Risk}} \\
\quad Mean     & \cellcolor{hb4}10.7 & \cellcolor{hb5}\textbf{17.7}  & \cellcolor{hb3}8.0  & 11.4 \\
\quad Max      & \cellcolor{hb4}60.8 & \cellcolor{hb5}\textbf{100.0} & \cellcolor{hb4}58.0 & 100.0 \\
\hline
\end{tabular}
\vspace{0.3em}
\footnotesize
\\
\textit{Note.} Color intensity encodes magnitude per row. Mean L3 and Max L3 denote the mean and maximum exploitability risk score under L3 verification, respectively.
\end{table}
Overall, these distributions reveal three cross-dimensional relationships. First, CWE family is associated with axial category: API-substitution vulnerabilities such as weak cryptography and deserialization map predominantly to Omission, logic-intensive vulnerabilities such as path traversal and XSS map to Inadequacy, and vulnerability types such as information disclosure and open redirect map to Introduction. This association indicates that vulnerability type helps characterize how silent failures manifest in the observed corpus. Second, each axial category was concentrated in a characteristic code location: Omission concentrates in API calls, Inadequacy in input handling, and Introduction in configuration artifacts. This separation is consistent with the taxonomy dimensions capturing distinct aspects of the failure phenomenon. Finally, severity does not follow a simple gradient across the three categories; instead, Inadequacy is associated with the highest exploitation risk despite representing attempted rather than absent defenses. 
 
\begin{tcolorbox}[colback=gray!5, colframe=gray!88, title=Key Findings from RQ3]
We found that different vulnerability contexts are associated with different types of silent failures in our corpus. API-substitution vulnerabilities are predominantly associated with Omission, logic-intensive vulnerabilities with Inadequacy, and vulnerabilities introduced during repair with Introduction. Among these categories, Inadequacy exhibits the highest estimated exploitability risk. These results indicate that CWE family and code location help characterize the types of silent failures observed in this study.
\end{tcolorbox}

\section{Discussion}
\label{sec:discussion}

This section discusses the key findings for each research question and outlines their implications for research and practice. It examines why patches that pass functional validation can remain insecure (RQ1), why current evaluation and review practices do not surface these cases (RQ2), and how the observed failure patterns can inform more targeted verification (RQ3).

\subsection{Security-Compliance Decoupling as a Distinct Failure Phenomenon (RQ1)}
\label{sec:disc_rq1}

\textbf{Silent failures as a structurally distinct class of defects.} The security-compliance decoupling we identified in our results (see Section~\ref{sec:rq1}) is structurally distinct from conventional software defects. In an ordinary bug, the code fails to produce expected behavior, and this failure is in principle detectable through functional testing. In a silent failure, by contrast, the code not only runs but produces the expected result; the deficiency lies in a non-functional dimension, namely the security property of the output, that the evaluation oracle was not designed to assess. This distinction matters because it implies that silent failures are unlikely to be eliminated by improving test coverage alone. Instead, they appear to require evaluation criteria that extend beyond functional correctness into security semantics.

This observation aligns with a long-standing finding in automated program repair, namely that plausible patches are not necessarily correct~\citep{qi2015analysis}. The cases we identify, however, go a step further: the patches are not merely plausible but also pass the available functional checks, yet remain insecure at the semantic level. As LLM-based agents become more capable of satisfying increasingly demanding test cases, this gap between functional adequacy and security correctness may widen rather than narrow.

\textbf{The two prominent open codes reflect different reasoning gaps.} At the axial level, the cases concentrate in a small part of the taxonomy: among the ten open codes, the two prominent codes, Insecure API Preserved (42) and Weak Mitigation (36), together account for nearly half of the corpus (78/170, 45.9\%). This concentration suggests that security-compliance decoupling is driven by a few recurring failure mechanisms rather than by many isolated factors.

Of the two, Insecure API Preserved is the more prevalent (42 cases) and illustrates the decoupling in its most direct form: the agent completes the functional task while leaving in place a dangerous API call that was already present in the input context. One plausible explanation is insecure pattern reproduction: the agent appears to treat the surrounding code as authoritative and edits around the insecure call rather than replacing it,
particularly when that call is not the direct subject of a failing test. As a result, the inherited insecure construct remains in the repaired code, even though the task itself calls for a security fix.

Weak Mitigation, by contrast, reflects a different dynamic, an attempted but incomplete defense. In our taxonomy, Weak Mitigation is the sole open code under Inadequacy, and among the recurring open codes it has the highest mean exploitability score (L3 = 17.7), which challenges the intuition that any defense is better than none. A plausible reading is that this gap lies between recognizing a vulnerability and implementing a robust defense: producing a partial defense requires the agent to recognize the vulnerability class, but not necessarily to reason about the conditions under which the defense actually holds. For example, the agent may recognize that an input should be constrained, yet implement a check that known bypass strategies can evade. Consequently, such a visible control can create an appearance of safety that the underlying code does not support.

The remaining open codes further indicate when the dominant mechanisms change. Missing Mitigation appears when the required security control must be inferred from task intent rather than substituted from the input, which is more demanding than replacing a visibly unsafe API. Deployment Context Risk and Unsafe Default Retained suggest that agents tend to treat deployment assumptions as peripheral rather than as part of the repair obligation, while Hardcoded Secret points to a related tendency to produce runnable examples rather than production-safe security artifacts. Missing Sanitization and Ambiguous Repair Intent appear in only a few cases and mark boundary conditions rather than recurring mechanisms. Taken together, these lower-frequency codes suggest that future agent design may need to distinguish API-substitution tasks from tasks that instead require new security reasoning, deployment-context reasoning, or secure-default generation.

These two dominant mechanisms are consistent with prior evidence on the structural limitations of LLMs. The tendency to preserve insecure context aligns with reported difficulties of agentic systems in reliably interpreting user intent~\citep{han2024llm,bansal2024challenges}: the observed behavior is consistent with satisfying the literal task while failing to address its broader security purpose. Likewise, the incompleteness of attempted defenses aligns with evidence that LLM-based vulnerability repair depends on context when applied to real-world vulnerabilities~\citep{pearce2023examining}, and with the long-standing automated program repair finding that a test-passing patch may mask rather than resolve the underlying defect~\citep{qi2015analysis}.

\textbf{Different decoupling modes may call for different remediation
strategies.}
The three axial categories appear to represent qualitatively different points at which the decoupling occurs. Omission involves a gap between understanding the task and selecting an appropriate API or control. Inadequacy involves a gap between identifying the security requirement and implementing a complete defense. Introduction involves a gap between
addressing the primary task and managing the security context of all generated artifacts. The heterogeneity of these failure modes is consistent with the MAST taxonomy, which likewise finds that failures in multi-agent systems fall into qualitatively distinct categories rather than a single defect type~\citep{cemri2026why}. In the verification literature, hybrid pipelines that combine LLM-based generation
with static analysis or formal methods~\citep{tihanyi2025vulnerability,dolcetti2025dual} similarly indicate that a single class of checks may not be sufficient for all deficiency types. If these distinctions hold, no single mitigation strategy would suffice: Omission may
respond to API substitution rules or secure-by-default libraries; Inadequacy may require deeper security reasoning or defense completeness verification; Introduction may require extending the verification boundary to all artifacts generated during the repair, rather than only the prompted target.

\textbf{\textit{Implications.}} Taken together, these results suggest that security deficiencies should not be treated as a monolithic category, and that verification pipelines may benefit from mode-specific checks matched to the expected failure profile of a given task. For researchers, the dominance of a single, mechanically simple code indicates
that a substantial portion of silent failures may be addressable through targeted, rule-based interventions, while the risk concentration in Weak Mitigation points to completeness-oriented verification, such as differential testing against known bypasses, as a priority for evaluation. For agent developers, safe-API substitution rules or curated secure-default mappings, applied before a patch is accepted, could remove a large share of the most common failures. For security engineers, agent-generated controls should be evaluated for completeness against known bypass techniques, not merely for presence, because a bypassable defense can lead downstream reviewers to assume that the issue has been closed and may therefore carry greater practical risk than an absent one.

\subsection{Why Current Assurance Pipelines Miss Silent Failures (RQ2)}
\label{sec:disc_rq2}

\textbf{Current evaluation paradigms may leave security-relevant gaps.}
Industry-standard benchmarks such as SWE-bench~\citep{jimenez2024swebench}
and HumanEval~\citep{chen2021evaluating} treat test-passing as the primary success criterion. Under these criteria, every confirmed case in our corpus would have been classified as a successful repair. This does not diminish the value of functional benchmarks as operational baselines, but it suggests that benchmark-reported success rates for security-sensitive tasks may overestimate actual repair quality. Complementing functional checks with security-oriented verification appears necessary to surface deficiencies that tests alone were not designed to capture. This concern is consistent with recent APR surveys, which note that current benchmarks offer few metrics for vulnerabilities introduced during repair~\citep{puvvadi2025coding}.

\textbf{Failures often originate early and survive to the final patch.} The propagation analysis suggests that reviewer failure is only one part of a broader chain-level problem. In many cases, the deficiency originated upstream and then passed through subsequent roles without security-focused re-evaluation. Although reviewer agents rarely intercepted confirmed deficiencies in our corpus, treating this solely as a reviewer limitation would understate the role of earlier pipeline stages, because the deficiency was already present before review, and later roles tended to assess task progress rather than reconsider the security adequacy of prior decisions. A plausible interpretation is that agent handoffs preserve functional intent more reliably than security intent: once a security control is omitted or weakened at design or implementation time, subsequent roles rarely reconstruct the missing requirement in our cases.

This lifecycle pattern relates closely to prior work on failure dynamics in multi-agent systems. Studies of cascading failures and problem drift describe how errors compound across interaction steps~\citep{bisconti2025beyond}, and attribution frameworks such as AgenTracer localize fault-inducing steps within agent trajectories~\citep{zhang2026agentracer}. These approaches, however, share a common limitation: they often assume an observable failure signal, such as an explicit error or a failing test, to define and trace the fault. Our propagation analysis, by contrast, suggests that similar cross-step dynamics can also arise when no such signal is present. This extends the study of multi-agent failure to latent deficiencies that remain invisible to functional evaluation.

\textbf{Convergence patterns are consistent with shared experimental factors, while failure profiles differ across architectures.}
Convergence is a common pattern in our corpus: 107 of the 170 confirmed failures fall within convergent cases, in which 26 tasks triggered comparable failures across three or more frameworks. Since the frameworks differ in orchestration and coordination architecture, this recurrence is consistent with shared factors across the experimental setup, including the base model, prompt content, task design, and recurring vulnerability patterns. This reading is further supported by the earlier observation that agents across the examined architectures tended to reproduce insecure APIs present in the task context. We do not, however, attribute the convergence to the model alone: because our design fixes the base model and varies the frameworks, it cannot separate model-level effects from shared factors such as prompt content, dataset characteristics, and recurring vulnerability patterns.

Architecture is nevertheless associated with differences in how these failures manifest. The orchestration of each system appears to be associated with how shared failure patterns surface in the generated artifacts: multi-agent systems, which involve more inter-agent interaction and additional artifact generation, exhibited more Introduction failures, whereas single-agent systems exhibited more Inadequacy failures, consistent with attempted but incomplete defenses.

This pattern extends prior evidence on the reliability of LLM-based repair. Earlier work reports that weaknesses recur across prompts and scenarios rather than as isolated, framework-specific errors~\citep{pearce2023examining}; our results carry this observation from individual models to agentic repair systems. They suggest that, as long as these recurring security limitations remain unaddressed, similar insecure behavior can recur even under different coordination mechanisms.

\textbf{\textit{Implications.}} These findings imply that security assurance for agentic repair should be evaluated and designed beyond functional pass rates. For benchmark designers, security-oriented criteria need to be reported alongside functional outcomes so that test-passing is not mistaken for successful security repair. For system designers, assurance mechanisms should move upstream: because silent failures can originate during planning, design, or implementation and persist through subsequent handoffs, tool-supported checks should be deployed where defects first arise rather than only at the final review stage. For researchers, the convergent subset identified in this study offers a useful basis for controlled experiments that vary base models under fixed architectures, or architectures under fixed models, to separate model-level vulnerabilities from architecture-level effects.

\subsection{Structured Failure Patterns as a Basis for Targeted Verification (RQ3)}
\label{sec:disc_rq3}

\textbf{Vulnerability family and code location jointly indicate the expected failure category.}
The observed association between CWE family and axial category suggests that vulnerability type may help anticipate the type of silent failure an agent is likely to produce. API-substitution vulnerabilities were predominantly associated with Omission; logic-intensive vulnerabilities with Inadequacy; and vulnerabilities introduced during repair with Introduction.

A parallel pattern appears at the level of code location: Omission is concentrated in API calls, Inadequacy in input-handling logic, and Introduction in configuration artifacts. A plausible explanation is that each category's gap surfaces where its work happens. An unsubstituted dangerous call remains visible at the call site; an incomplete defense appears in the input-handling code intended to enforce it; and a newly introduced weakness often appears in configuration or auxiliary artifacts generated by the pipeline. Both associations are consistent with prior evidence that LLM-based repair performs more reliably on well-specified, pattern-like fixes than on vulnerabilities whose correct handling depends on context~\citep{pearce2023examining}.

\textbf{Multi-agent pipelines can introduce vulnerabilities outside the originally targeted patch scope.} The co-occurrence of multi-agent architecture, the Introduction category, and configuration-level locations is one of the more distinctive patterns in our data. During execution, multi-agent frameworks may emit scaffolding beyond the
requested patch, such as build files, container definitions, runtime settings, and other auxiliary artifacts. These artifacts can contain weaknesses that were absent from the original task. Two factors make such weaknesses difficult to detect: they are not the explicit target of the prompt, so functional tests may not exercise them, and pipeline roles are
typically not assigned to review every generated artifact for security. Existing failure taxonomies for multi-agent
systems~\citep{cemri2026why} and vulnerability classifications such as CWE help characterize failures or vulnerabilities once they are observed, but they do not address the possibility that the analysis boundary itself may be too narrow. Our results extend this picture by showing that the set of security-relevant artifacts can grow with pipeline complexity.

\textbf{Silent failures are not confined to low-severity issues.} The severity distribution shows that security-compliance decoupling is not limited to minor weaknesses: 22 of the 170 confirmed cases were rated as critical severity, yet they remained undetected by functional testing in the same way as lower-severity cases. This result indicates that functional success does not provide a severity-sensitive assurance signal, because a patch may pass all available tests while still preserving or introducing high-impact vulnerabilities. This observation aligns with arguments in the verification literature for security-oriented test suites that assess repairs beyond functional correctness~\citep{kuzmina2025spring,kaniewski2025systematic}.

\textbf{\textit{Implications.}} For practitioners, the family and location associations indicate that verification effort can be routed rather than applied uniformly: rule-based API checks at call sites for API-substitution families, deeper logic analysis on input-handling code for logic-intensive families, and scanning that covers generated configuration and auxiliary artifacts for newly introduced vulnerability types. Moreover, because multi-agent pipelines can introduce weaknesses outside the prompted target, vulnerability analysis of agent output should enumerate and scan every
artifact the pipeline produces, rather than only the file or component named in the task prompt. For reviewers, the presence of critical-severity cases among silent failures argues against taking a passing test suite as evidence of a secure repair or of acceptable residual risk; independent security verification should therefore precede any decision to trust an agent-generated patch. For researchers, how the set of at-risk artifacts scales with pipeline complexity remains unquantified and could be measured directly or through iterative re-analysis of the patched code.

\section{Threats to Validity}
\label{sec:threats}

We organize threats to validity following the standard validity categories in empirical software engineering~\citep{wohlin2012experimentation}.

\subsection{Construct Validity}
 
Our operational definition identifies silent failures as patches that satisfy both syntactic and functional verification yet are flagged by static security analysis or exploitability assessment. This definition  focuses on security-relevant silent failures. Silent failures involving performance, privacy, maintainability, usability, reliability under load, or other non-security qualities are outside the scope of this study, and our counts should not be read as estimates of failures in those dimensions. Similarly, security properties requiring dynamic analysis, runtime monitoring, environmental configuration, or domain-specific knowledge may not be captured by our verification framework.
 
Our taxonomy was derived from all 170 confirmed cases, which include 39 (22.9\%)
classified as Partial alongside 131 (77.1\%) true positives. Including Partial cases may affect precision, because a Partial verdict reflects less direct evidence of exploitability than a true-positive verdict. As a sensitivity check, we re-examined the taxonomy using only the 131 true positives. All three axial categories and seven of the ten open codes remain represented in this subset. The three omitted codes occurred only as Partial cases: Deployment Context Risk (9 cases), Ambiguous Repair Intent (1 case), and Vulnerable Test Code (1 case). This is expected, since their exploitability could not be confirmed from static evidence alone, which is the same property that placed them in the Partial category. Although this restriction reduced counts across all three categories, the main taxonomy structure remained stable.
 
The distinction between Omission and Inadequacy depends on whether the agent made any attempt at mitigation. Because the two categories are reported separately, misclassifying boundary cases would shift their relative sizes. We used any observable mitigation attempt in the trace as the deciding criterion and resolved the small number of boundary cases through discussion between coders.

\subsection{Internal Validity}
 
The L2 verification level relies on Bandit, which is subject to both false positives and false negatives. False positives were addressed through manual review, which excluded 82 candidates from the confirmed corpus. False negatives are inherently difficult to quantify; the confirmed corpus should be interpreted as a conservative estimate of the silent failures detectable by our multi-level verification framework (see Section~\ref{sec:verification}), not as an estimate of all possible silent failures in agentic code repair.

Attributing a silent failure to a specific agent role and pipeline stage requires interpretive judgment, particularly where role boundaries are ambiguous in multi-agent traces. Because originator attribution was derived from qualitative trace analysis, alternative interpretations of complex agent interactions could lead to different role assignments. This threat may influence the reported originator distributions and the propagation-chain frequencies. To mitigate this threat, we relied on observable trace evidence, predefined coding rules, and consensus-based discussion of ambiguous cases.

A related threat concerns the classification of propagation patterns. Patterns such as planning-stage omission, design-originated omission, design-originated introduction, and implementation-originated failure were derived through qualitative analysis of the traces, and some cases could reasonably be assigned to more than one pattern, especially when underspecified requirements and implementation-level omissions occurred in the same trace. Because the frequency of each pattern feeds into our findings, ambiguous assignments could shift the reported pattern distribution. To reduce this threat, we recorded the trace evidence behind each assignment, compared each newly coded case against earlier cases in the same pattern to limit category drift, and discussed boundary cases jointly before assigning a single final pattern.

The L3 exploitability score is a verification level separate from L2. It relies on predefined signal weights, bounded safe-practice discounts, and a fixed candidate threshold (see Section~\ref{sec:verification}). The fixed threshold therefore affects which cases are judged as silent failure candidates. For example, a true silent failure scoring below the threshold would not enter manual review and would be missed. The score magnitude, in turn, affects the reported exploitability patterns, so the weighting and discount choices, together with any imprecision in the pattern matching of sources and sinks, contribute to these patterns and to the results that build on them. 

To reduce this threat, every candidate above the threshold was manually reviewed before confirmation, so the threshold acts as a screening sensitivity parameter rather than a final classification boundary, and over-inclusion at screening is removed by review. We also applied the same scoring procedure uniformly across all frameworks, so comparative claims rely on the ordering of scores rather than their absolute values.
 
\subsection{External Validity}
 
Both datasets are Python-focused and may not represent the full diversity of security-sensitive repair tasks encountered in practice. Tasks involving multi-file repositories, complex dependency chains, or language-specific security idioms are underrepresented. We therefore limit our claims to Python repair tasks, and do not generalize to other languages or to repository-scale repair.
 
The seven frameworks (four multi-agent, three single-agent) span a range of coordination architectures, but the landscape of agent systems is evolving rapidly. Our sample of seven frameworks may therefore not generalize to proprietary systems. Despite this architectural diversity, because all frameworks were instantiated using GPT-4o-mini, some observed failure patterns may reflect characteristics of the underlying model rather than the framework architecture itself. The cross-framework convergence we observed sharpens this concern: it is consistent with shared model-related factors, but it may also be influenced by prompt templates, the characteristics of the two benchmarks, the structure of the repair tasks, recurring vulnerability patterns, or other experimental factors. We therefore treat the source of this convergence as an open question rather than evidence of a single cause, and we cannot fully separate model-specific effects from framework-specific ones. Future studies should replicate the analysis using additional foundation models.

\subsection{Conclusion Validity}

Conclusion validity concerns whether our results provide support for the conclusions we draw. In this study, the analysis is primarily qualitative and descriptive, so the conclusions are based on repeated patterns and frequency distributions within the confirmed corpus. Several design choices support these conclusions. First, all cases were processed through the same multi-level verification and manual review procedure before entering the confirmed corpus. Second, each confirmed case was assigned to one dominant axial category, which makes the category-level distributions mutually exclusive. Third, the coding criteria were applied consistently across iterations, with boundary cases discussed before final assignment. Lastly, comparisons across frameworks are based on executions conducted under a shared model configuration, timeout, and verification pipeline. These choices reduce threats to conclusion validity. The reported counts are still corpus-dependent, and small differences between adjacent categories, frameworks, or open codes should not be interpreted as stable rankings. We therefore report raw counts together with percentages, describe the results as patterns observed in our corpus, and avoid drawing causal or population-level conclusions from descriptive differences alone.

\subsection{Reliability}
 
Qualitative coding was performed by the first author across all three iterations. A random sample of 35 cases was independently coded by another author, yielding a composite Cohen's $\kappa$ of 0.71. This composite requires agreement on both the manual review label (TP, PARTIAL, or FP) and the axial category (Omission, Inadequacy, or Introduction), with
per-dimension values of 0.63 and 0.66 respectively; the procedure is detailed in
Section~\ref{sec:reliability}. Disagreements were resolved through structured consensus discussion.
 
The verification pipeline is deterministic given the same input. Framework execution involves stochastic LLM inference, meaning that exact trace reproduction is not guaranteed.

\section{Related Work}
\label{sec:related}
To reduce the risk of overlooking relevant studies, we searched seven major digital libraries: Google Scholar, IEEE Xplore, the ACM Digital Library, ScienceDirect, Scopus, Web of Science, and SpringerLink. To the best of our knowledge, we found no prior work that traces how patches that pass available functional checks but remain insecure originate in agent roles and propagate through agentic repair pipelines. We discuss the existing research in four categories: (i) benchmark-based evaluation of LLM agents (Section~\ref{sec:related_benchmark}), (ii) failure analysis in multi-agent systems (Section~\ref{sec:related_failure}), (iii) the plausible-versus-correct distinction in APR (Section~\ref{sec:related_apr}), and (iv) security verification of generated repairs (Section~\ref{sec:related_security}). A concluding summary and comparative analysis (Section~\ref{sec:related_summary}) position the present study at the intersection of these lines and justify its scope and contributions.

\subsection{Benchmark-Based Evaluation of LLM Agents}
\label{sec:related_benchmark}

Evaluation of LLM-based agentic systems is often driven by benchmarks. SWE-bench has become a widely used benchmark for assessing automated repair, with related efforts adopting similar outcome-focused metrics that measure whether a generated patch passes existing unit tests~\citep{jimenez2024swebench,peng2024survey,wang2024battleagentbench}. These metrics provide a useful operational baseline, but  they treat a passing result as success without assessing whether the generated patch satisfies the security intent of the repair task. As a result, a patch that passes existing tests may still preserve the vulnerable behavior in the original code. Outcome-level evaluations therefore offer limited insight into failure mechanisms or non-functional properties of the output. Beyond these outcome metrics, researchers deploying agentic systems in increasingly complex workflows have begun examining broader limitations such as misalignment, coordination breakdowns, and difficulty in reliably interpreting user intent~\citep{han2024llm,hammond2025multi,bansal2024challenges}. These perspectives have helped frame the challenge landscape but remain primarily qualitative, with limited empirical analysis of specific failure patterns.

\subsection{Failure Analysis in Multi-Agent Systems}
\label{sec:related_failure}

A growing body of work has moved toward empirical failure analysis in multi-agent systems. Agentic systems vary in architecture, including single-agent and multi-agent designs with different role structures and coordination patterns, which may influence where failures appear~\citep{Masterman2024TheLO}. The MAST taxonomy provides a systematic categorization of observable failure modes~\citep{cemri2026why}, while complementary work has formalized dynamics such as cascading failures and problem drift at the system level~\citep{bisconti2025beyond,reid2025risk,becker2024multi}. These efforts characterize failures that surface as errors or task non-completion, but they do not address failures that give no such outward indication. Attribution frameworks have advanced toward more granular diagnosis: Who\&When and AgenTracer introduced fine-grained annotation and counterfactual replay to identify fault-inducing steps within complex trajectories~\citep{zhang2025which, zhang2026agentracer}. Their localization, however, depends on an observable failure signal to define the fault being traced. Beyond diagnosis, resilience mechanisms such as challenger roles and courtroom-inspired inter-agent debate have been proposed to improve reliability~\citep{huang2024resilience,widyasari2025let}. These approaches demonstrate the potential of role-based specialization, though they rely predominantly on explicit failure signals such as crashes, error messages, or task non-completion. Failures that do not produce such signals remain largely outside their scope. This gap motivates the present study.

\begin{table*}[t]
\centering
\scriptsize
\setlength{\tabcolsep}{3.5pt}
\renewcommand{\arraystretch}{1.25}
\begin{threeparttable}
\caption{Positioning of representative prior work relative to the present study. 
Y = explicitly addressed; P = partially addressed; -- = not a primary focus.}
\label{tab:related_positioning}
\begin{tabular}{@{}p{0.20\textwidth}
                cccccc
                p{0.28\textwidth}@{}}
\toprule
\textbf{Representative work} 
& \textbf{Bench.} 
& \textbf{Taxon.} 
& \textbf{Prop.} 
& \textbf{Role attr.} 
& \textbf{Sec. verif.} 
& \textbf{Silent fail.} 
& \textbf{Main gap relative to this study} \\
\midrule
Outcome benchmarks 
(e.g., SWE-bench)~\citep{jimenez2024swebench}
& Y & -- & -- & -- & -- & -- 
& Evaluate whether patches pass existing tests, but do not assess security-semantic correctness or explain latent security failures. \\
\midrule
Multi-agent failure taxonomies 
(e.g., MAST)~\citep{cemri2026why}
& P & Y & P & -- & -- & -- 
& Characterize observable multi-agent failures, but do not cover failures that leave no external signal such as a crash, error, or failed task. \\
\midrule
Trajectory attribution frameworks (e.g., AgenTracer)~\citep{zhang2025which, zhang2026agentracer}
& -- & -- & Y & Y & -- & -- 
& Localize fault-inducing steps in agent trajectories, but require an observable failure signal to define what should be traced. \\
\midrule
APR plausible-versus-correct 
studies~\citep{qi2015analysis,wu2023effective,puvvadi2025coding}
& P & P & -- & -- & P & P
& Establish that test-passing patches may be incorrect, but do not analyze security-specific silent failures in role-based agentic repair pipelines. \\
\midrule
LLM security repair and 
verification~\citep{pearce2023examining,tihanyi2025vulnerability,dolcetti2026helping,mahmud2025systematic,low2024repairing}
& P & -- & -- & -- & Y & P
& Assess the security of generated repairs at the artifact level, but do not explain how deficiencies originate, propagate, or evade review across agent roles. \\
\midrule
\textbf{This study}
& Y & Y & Y & Y & Y & Y
& Provides an empirical taxonomy of silent failures, analyzes their role-level origins and propagation chains, and verifies security deficiencies that remain hidden after syntactic and functional success. \\
\bottomrule
\end{tabular}
\begin{tablenotes}[flushleft]
\footnotesize
\item \textit{Dimensions.} \textbf{Bench.}: benchmark-based evaluation of
repair outcomes; \textbf{Taxon.}: classification of failure types;
\textbf{Prop.}: how failures propagate across pipeline stages;
\textbf{Role attr.}: attribution of a failure to a specific agent role;
\textbf{Sec.\ verif.}: verification of the security of generated code beyond
functional correctness; \textbf{Silent fail.}: coverage of failures that pass
syntactic and functional checks yet retain or introduce a vulnerability.
\end{tablenotes}
\end{threeparttable}
\end{table*}

\subsection{APR and the Plausible-versus-Correct Distinction}
\label{sec:related_apr}

While the preceding work examines failure dynamics within agentic pipelines, the failures themselves ultimately materialize in repair artifacts whose quality must also be assessed against established standards in Automated Program Repair (APR) and software security. A central concern in APR is the distinction between plausible and correct patches. A plausible patch passes all tests in the validation suite yet may not correct the defect, and can eliminate desirable functionality or even introduce new security vulnerabilities~\citep{qi2015analysis}. Recent APR surveys characterize the field as having evolved from template-based and constraint-based repair methods to machine learning, deep learning, and more recently large-language-model-based approaches~\citep{dolcetti2026helping,puvvadi2025coding}. In this evaluation-driven setting, the security of the resulting patches has received comparatively limited attention, and current benchmarks offer few metrics for vulnerabilities introduced during repair~\citep{puvvadi2025coding}. 

This distinction is the conceptual basis for the silent failures we study, narrowed to the security setting. In APR, the plausibility gap concerns whether a test-passing patch actually fixes the target defect. In our setting, the gap concerns a security property that the available functional tests do not observe: the generated patch passes syntactic and functional validation, but still leaves the target vulnerability unaddressed or introduces a new one. This security-oriented reading of plausibility has begun to appear in repair evaluations: when a patch is counted as plausibly fixed once it passes the project test cases, neural and LLM-based repair models still fail to address many vulnerability types, including cryptographic and request-handling weaknesses~\citep{wu2023effective}. Such cases show that test passage and security correctness can diverge even when the repair target is a known vulnerability. This limitation is also reflected in LLM-based vulnerability repair, where prior evaluations show that models perform reasonably well in controlled settings but remain unreliable when applied to real-world vulnerabilities~\citep{pearce2023examining}. However, these studies primarily evaluate individual patches generated by off-the-shelf models and do not examine how such deficiencies arise and propagate across the roles of a multi-agent repair pipeline. Our study addresses this gap.

\subsection{Security Verification of Generated Repairs}
\label{sec:related_security}

A separate line of work addresses this gap through security-oriented verification. Established security test suites have been used for assessment beyond functional correctness~\citep{kuzmina2025spring, kaniewski2025systematic}, while hybrid pipelines combine LLM-based generation with formal verification or static analysis~\citep{tihanyi2025vulnerability, dolcetti2025dual}. In industrial practice, such verification is increasingly embedded in CI/CD pipelines~\citep{meliala2024integrating}. Even where such verification is applied, however, generated fixes can satisfy the surface checks while leaving the underlying weakness in place; in infrastructure-as-code repair, for instance, a notable share of LLM-generated fixes passed configuration checks yet did not resolve the security issue and escaped scanner detection~\citep{low2024repairing}. This indicates that artifact-level verification reduces, but does not remove, the risk of security deficiencies that survive functional and tool-based checks. Adversarial evaluation sharpens this point: under systematic bypass testing, AI-generated patches were found to rely on minimalistic
input validation and to remain exploitable through simple attack techniques,
in contrast to the more structural corrections in human-written patches~\citep{mahmud2025systematic}. These approaches assess the security of a patch as a finished artifact, but they do not model how the deficiency arose within the repair process, nor which role in a multi-agent pipeline produced it. As a result, security verification and multi-agent repair remain insufficiently connected, with no account of how a vulnerability passes through agent interactions and evades review.

\subsection{Conclusive Summary}
\label{sec:related_summary}

These complementary lines of work have each advanced substantially but remain only partially connected. Benchmark studies treat test success as the primary quality indicator and do not examine the security semantics of a passing patch. Failure-analysis studies focus on breakdowns that surface as explicit signals and do not account for outputs that appear correct yet contain latent vulnerabilities. APR research establishes the distinction between plausible and correct patches, and recent work shows that LLM-based security repair remains unreliable, but it analyzes individual patches rather than the roles that produce them. Security-verification studies assess patch quality at the artifact level without modeling how the internal role interactions of a multi-agent pipeline contributed to the deficiency. What remains unaddressed is the intersection of these concerns: how patches that pass functional evaluation yet preserve or introduce exploitable vulnerabilities originate within agent role structures, propagate through pipeline stages, and evade review mechanisms. The present study targets this intersection. Table~\ref{tab:related_positioning} positions our study against representative prior work along these dimensions and illustrates that the representative prior work considered here does not jointly address agent-role attribution, propagation, security verification, and silent-failure coverage.

\section{Conclusion}
\label{sec:conclusion}

We present an empirical study of silent failures in LLM-based agentic code repair. To investigate this phenomenon, we analyzed 170 confirmed silent failure cases identified from 1,030 valid execution traces produced by seven agent frameworks across two security-focused datasets. The primary contribution of this study is an empirical characterization of these failures along three dimensions: a three-part taxonomy (Omission, Inadequacy, and Introduction) grounded in confirmed cases, an account of how they propagate through agent pipelines, and an analysis of how they distribute across frameworks, vulnerability families, and code locations. Based on the analysis of the results, the key findings of this study are as follows:
\begin{itemize}
\item Our results show that silent failures should be treated as a separate form of security-compliance decoupling rather than as ordinary functional defects alone. The confirmed cases support a three-part structure consisting of Omission, Inadequacy, and Introduction.

\item We also observed that current assurance pipelines centered on test passage and LLM-based review were insufficient to detect these failures. At the same time, the recurrence of similar insecure solutions across frameworks is consistent with shared factors in the experimental setup, including the base model, prompts, tasks, and vulnerability patterns, alongside differences in how failures appear across agent architectures.
\item Our results show that silent failures exhibit recurring patterns across vulnerability families, code locations, and generated artifact types in our corpus. These patterns suggest that security assurance for agentic repair should move toward targeted verification that covers the full repair scope rather than general post hoc inspection.
\end{itemize}

These findings have implications for both research and practice. For research, they suggest that success in agentic repair should no longer be defined solely by whether code compiles and passes tests. Future benchmarks and evaluation protocols should assess both functional validation and security-semantic correctness, and should examine base-model effects and orchestration effects through controlled experimental designs. For practice, they indicate the need to extend verification beyond the primary patched file to all generated artifacts and to augment reviewer roles with tool-backed security checks.

Future work should replicate these findings across models, languages, and tool-supported agent designs, and test whether targeted verification strategies meaningfully reduce silent failure rates in deployment settings. Another direction is to examine agent execution traces to better explain how these silent failures arise, including where security considerations are weakened or dropped during the repair pipeline. This line of work may help shift assurance from post hoc detection toward earlier intervention before patches are accepted.

\printcredits
\section*{Acknowledgments}
This research is funded as part of Finland's Ministry of Education and Culture's Doctoral Education Pilot under Decision No. VN/3137/2024-OKM-6 (The Finnish Doctoral Program Network in Artificial Intelligence, AI-DOC).

\section*{Data availability}
Link to our dataset is in the reference \citep{Bai2026dataset}.

\section*{Declaration of AI Assistance}
During the preparation of this work, the author(s) used ChatGPT to refine grammar, improve sentence structure, and resolve formatting issues. After utilizing this tool, the author(s) thoroughly reviewed and edited the content as needed, taking full responsibility for the final publication.

\bibliographystyle{elsarticle-num}

\bibliography{cas-refs}

\end{document}